\documentclass[mathleft]{an}
\usepackage{graphicx}
\usepackage{times}
\usepackage{subfigure}
\usepackage{SIunits}

\begin{document}
\sloppy

\Pagespan{789}{}
\Yearpublication{2006}%
\Yearsubmission{2005}%
\Month{11}%
\Volume{999}%
\Issue{88}%

\title{Challenges in optics for Extremely Large Telescope instrumentation}

\author{P.~Span\`o\inst{1}\fnmsep\thanks{Corresponding author:
  \email{spano@oa-brera.inaf.it}}, F.M.~Zerbi\inst{1},
        C.J.~Norrie\inst{2}, C.R.~Cunningham\inst{2},
        K.G.~Strassmeier\inst{3},
        A.~Bianco\inst{1}, P.A.~Blanche\inst{4},
        M.~Bougoin\inst{5}, M.~Ghigo\inst{1},
        P.~Hartmann\inst{6}, L.~Zago\inst{7},
        E.~Atad-Ettedgui\inst{2}, B.~Delabre\inst{8}, H.~Dekker\inst{8},
        M.~Melozzi\inst{9}, B.~Sn\"yders\inst{10}, R.~Takke\inst{11},
        \and D.D.~Walker\inst{12,13}
        }
\titlerunning{Challenges in optics for ELT instrumentation}
\authorrunning{P. Span\`o et al.}
\institute{INAF-Osservatorio Astronomico di Brera, Via Bianchi 46, I-23807 Merate (LC), Italy
\and
UK Astronomy Technology Centre, Edinburgh, Blackford Hill, EH9 3HJ, UK
\and
Astrophysikalisches Institut Potsdam, An der Sternwarte 16, D-14482 Potsdam, Germany
\and
Centre Spatial de Li\`ege-ATHOL, Avenue du Pr\'e-Aily, B-4031 Angleur, Belgium
\and
BOOSTEC, Zone Industrielle, F-65460 Bazet, France
\and
Schott AG, Hattenbergstr. 10, D-55122 Mainz, Germany
\and
CSEM SA, Rue Jaquet-Droz 1, P.O.-Box CH-2007 Neuch\^atel, Switzerland
\and
European Southern Observatory, Karl-Schwarzschild-Str. 2, D-85748 Garching b. Muenchen, Germany
\and
Galileo Avionica, Via A. Einstein 35, I-50013 Campi Bisenzio (FI), Italy
\and
TNO TPD, P.O. Box 155, NL-2600 AD Delft, The Netherlands
\and
Heraeus Quarzglas GmbH, Quarzstr. 8, D-63450 Hanau, Germany
\and
Ultra Precision Surfaces Lab., OpTIC Technium, F.W. Morgan, St Asaph Business Park, N. Wales, LL17 0JD, UK
\and
Zeeko Ltd, 4 Vulcan Court, Hermitage Industrial Estate, Coalville, Leichestershire,
England, LE67 3FW, UK
}

\received{3 March 200} \accepted{27 March 2006} \publonline{later}


\abstract{We describe and summarize the optical challenges for
future instrumentation for Extremely Large Telescopes (ELTs).
Knowing the complex instrumental requirements is crucial for the
successful design of 30-60m aperture telescopes. After all, the
success of ELTs will heavily rely on its instrumentation and this,
in turn, will depend on the ability to produce large and
ultra-precise optical components like light-weight mirrors, aspheric
lenses, segmented filters, and large gratings. New materials and
manufacturing processes are currently under study, both at research
institutes and in industry. In the present paper, we report on its
progress with particular emphasize on volume-phase-holographic
gratings, photochromic materials, sintered silicon-carbide mirrors,
ion-beam figuring, ultra-precision surfaces, and free-form optics.
All are promising technologies opening new degrees of freedom to
optical designers. New optronic-mechanical systems will enable
efficient use of the very large focal planes. We also provide
exploratory descriptions of ``old'' and ``new'' optical technologies
together with suggestions to instrument designers to overcome some
of the challenges placed by ELT instrumentation.}

\maketitle


\section{Introduction}

Extremely Large Telescopes (ELTs, ground-based telesco\-pes between
20m and 100m in diameter, i.e. substantially larger than any
currently operating facility) are desired by Astronomers mainly
because of the urgent need of spectroscopic support for future space
facilities such as the James Webb Space Telescope (JWST).
Instruments on-board JWST will reach magnitudes that require much
larger collecting areas to perform spectroscopic follow-up than any
telescope currently operating. Even today the Advanced Camera for
Surveys on the Hubble Space Telescope has already produced images of
a depth far exceeding the ability of current telescopes to acquire
useful spectra, with the exception of objects having strong emission
lines.

Apart from space facilities follow-up, ELTs will be breakthrough
astronomical facilities in their own right, able to tackle problems
such as  the nature of first-light objects, the evolution of large
scale structures in the early universe and of the chemical
composition of the intergalactic medium, the assembly of galaxies at
high red-shifts and the traces of this process today, the formation
of stars and planetary systems, the detection and characterization
of extra-solar planets, possibly down to terrestrial sizes, and
detailed synoptic studies of objects in our own solar system.
Detailed science cases for ELTs have been published, amongst others,
by Hawarden et al. (2003).

Europe is putting a lot of effort, both at research institutions and
at industry level, to pursue a feasible configuration for an ELT.
Two preliminary designs exist and are currently debated in the
community: the ESO 100 or 60-m OWL (OverWhelmingly Large) telescope
(Dierickx et al. 2003) and the 50-m aplanatic Gregorian named Euro50
(Andersen et al. 2003). On the American side three original
proposals for extremely large telescopes exist; the GSMT (Giant
Segmented Mirror Telescope), the Canadian VLOT (Very Large Optical
Telecope) and the CELT (California Extremely Large Telescope). The
latter two were recently merged into a single TMT (Thirty Meter
Telescope) project.

All these facilities are expected to be operated in seeing-limited
conditions and in diffraction-limited conditions (Strehl 0.2-0.5
over field of view of arcminutes via MCAO correction). In
seeing-limited conditions a telescope of diameter $D$ [metres] and
focal ratio $F_{\#}$ , a point source with FWHM of the image PSF
$\theta$ [arcsec], is recorded with a width in [microns] given by
\begin{equation}
  \label{eq1}
N_p=4.878F_\#\theta D
\end{equation}
where the width is indicated as the dimension of the individual
pixel $p$ multiplied by the number of pixels $N$. Although the
telescope's optical design is under study, it is very unlikely that
a solution with $F_\#<1$  will be found. In the case of
seeing-limited applications and assuming the limit case $F_\#=1$  ,
a median good-site seeing of  0".6, observed with a 100-m telescope
would produce a 293$\mu$m spot.  This has two immediate
consequences: a) The focal plane of ELTs will be very large in size,
b) current detector pixel dimensions (10-20$\mu$m), if used in this
context, provide heavily over-sampled images. One could accept such
an over-sampling if it were not for the problems of cost and
procurement. One could also encourage the development of larger
pixel detectors. Since read-out noise is generally proportional to
pixel area, there is a price to pay in terms of sensitivity,
specially in detector limited applications such as high resolution
spectroscopy. More importantly, the development and production costs
of such devices would be prohibitive.

The problem of \'etendue, driving the plate scale on the telescope
focal plane, can be solved by dividing the pupil between several
instruments. Six or 7 sub-pupils of 30m each can be obtained out of
a single 100-m pupil at the price of a limited waste of aperture.
Each of the sub-pupils would be identical to a stand-alone facility
of the correspondent size. If one wishes to fully exploit the depth
reachable with a 100-m pupil one will have to manufacture 6-7
identical instruments for a 30-m pupil and then coadd the signals.
On the contrary, if the pupil division system is specially designed
for this purpose, each of the subpupil could feed instruments with
different characteristics allowing one to cover simultaneously,
albeit at a more limited depth, a larger set of physical parameters,
e.g. different polarizations, wavelength, etc.. The above schemes
require to plan a limited series production of instruments
traditionally conceived as single units, assessing the problem of
high reproducibility of identical components specially in optics,
mechanics, detectors, etc..

Even in the divided-pupil option, the extreme focal ratios which are
required for instruments operating well away from the telescope
diffraction limit raise the issue of weight. Traditional optical
components with strong curvatures are invariably massive for their
size. In addition, any intermediate optical stage is likely to be an
order of magnitude slower so that FoV and optical elements will have
to be large, typically 1m for a 30-m telescope (or sub-pupil) and 3m
for a 100-m facility. Apart from manufacturing challenges, this
implies physically large instruments which could not always be
compatible with the ELT mechanical structure.

Most ELTs are designed to work with Adaptive Optics systems. Much
effort is currently devoted to the development of wide-field
high-performance AO systems capable of delivering Strehl ratios of
20-50\% over fields-of-view of arcminutes instead of arcseconds.
There are various technical approaches aimed to deliver the same
results, including MCAO (Multi-Conjugated Adaptive Optics). A 30-m
telescope with a near diffraction-limited FoV of about 2 arcmin at a
wavelength of 1~$\mu$m yields about 3$\times$108 resolution elements
each of about $7\,10^{-3}$ arcsec in size. A billion pixel detector
is needed to sample properly such a small area, to be compared with
the 4k$\times$4k currently available on the market for these
wavelengths. For a 100-m telescope the situation is about one order
of magnitude more difficult. The only solution in this case is a
system to pick-off sub-fields for full resolution observing, i.e.
divide the focal plane. In the diffraction-limited regime the
pick-off is determined by the plate-scale achievements and is common
to any kind of instrument willing to exploit the full capability of
MCAO.

In order to address the technology improvements needed to face the
phase of ELT-instrument design and manufacture, a workshop was
jointly organized by the Italian National Institute of Astrophysics
(INAF) and the UK Astronomy Technology Centre (UKATC), under the
auspices of the OPTICON Key Technology Network. This workshop joined
together European institutes and industries for a bra\-instorming
discussion about problems and possible solutions. Its key issues,
main summaries and preliminary conclusions are presented in this
paper.


\section{Large optical components}

\subsection{Summary of requirements for large optics}

The Extremely Large Telescopes (TMT, LSST, European ELT) and
their instrumentation will require large optical components:
mirrors, lenses, filters, and gratings, with very stringent
specifications. We review the requirements in large optical
components featured in some of the ELT instruments. This includes
the specifications of the optical materials in terms of homogeneity
and stability in index of refraction, polishing errors and
anti-reflection coatings. Issues concerning the mounting of these
components at various environmental conditions must be addressed.
For each effect we will explore implications and possible new areas
of technology development.

\subsubsection{Optical blank specifications}

\emph{Clear Aperture Diameter.} For BK7, Fused Silica, and FK5 we
will possibly need 1.2m to 2m. These will be used mainly for
windows, field correctors, ADCs. In the IR,  materials such as
BaF$_2$, CaF$_2$, ZnS, ZnSe are very popular in cameras and
collimators. Can we break the boundary of the limited size we can
order from optical industry?  At  present, we can find these
materials at a diameter of 200mm to 500mm. In the sub-mm (250, 450
and 850$\mu$m wavelength), the only transparent materials used are
in polystyrene or plastic materials. We will need windows of 1 m,
filters of around 300 to 500mm, dichroics around 200mm diameter.

\emph{Homogeneity.} Some of the materials will be required to have
10$^{-7}$ for 1 m diameter and 60mm thickness.  At present 10$^{-6}$
can possibly be achieved for special homogeneous materials. The
measurement of homogeneity for large pieces is an issue: at present
only 500mm can be measured (``Schott Direct Measuring
Interferometer'').

\emph{Striae:} $<$10 nm wavefront distortion.

\emph{Bubbles and inclusions:} $<$0.05mm to 1mm or free of it.

\emph{Stress birefringence:} $<$4 nm/cm.

\emph{Index of refraction/dispersion.} More accurate measurements of
refractive indices ($n$) are required at several wavelengths and
cryogenic temperatures. The data of  $dn/dT$ and  $dn/d\lambda$ of a
large number of materials is not known. A serious campaign is needed
to be able to measure all these parameters.

\subsubsection{Specification of optical surfaces}

The processes of manufacturing such as rough grinding, fine
grinding, rough polishing, fine polishing using materials such as
pitches, diamond tools, etc. are well known in optical industries.
The main issues are the tougher and tougher requirements on these
processes to achieve diffraction limited optical instruments. Edge
effects are very important when manufacturing a segment in ELT.  All
these processes are relatively slow and expensive. New and
innovative processes such as replication, MRF (magneto-rheolo\-gical
fluid) or others could make a revolution in the optics industry of
large components. Detailed specifications are:

\emph{Radius of curvature:} From few millimeters to 100m. Very fast
f-ratios (f/0.8).

\emph{Asphericity or deviation from spheres:} From a few microns to
10mm.

\emph{Form errors:} Defined by quantities such as low spatial
frequency, medium spatial frequency, high spatial frequency surface
errors. They are expressed in terms of  peak-to-valley (nm) , RMS
(nm), Zernike coefficients, and power spectral density (PSD in
nm$^2$/m$^2$). The current achieved values are 20nm RMS for large
diameters. Requirements for specific applications are 0.3nm RMS
static and 0.075nm RMS differential.

\emph{Microroughness:} The very high spatial frequency surface
errors such as the microroughness and expressed in terms of RMS (nm)
and measures the BRDF or scattered light in the surface. Current
values 1 nm RMS.

\emph{Cosmetic status of the surface:} Dust or other contaminates,
water, anti-reflection or reflective coatings, scratches and digs.

\subsubsection{ELT instrument preliminary optical requirements}

\emph{High-dynamic imagers:} For direct imaging of Earth-type
planets one needs to obtain a contrast of 2\,10$^{-10}$ relative to
the host star. Therefore, one needs to achieve (after correction
with adaptive optics):
      \begin{itemize}
      \item A total wavefront error RMS$<$0.3 nm static across 1 to 150
      cycles/pupil,
      \item a differential aberration before the
      coronograph less than 0.075 nm,
      \item a differential aberration after the coronograph of $<$1nm.
      \end{itemize}
\noindent These numbers are very small and not achievable today.
There may be two possible solutions:
      \begin{itemize}
      \item Improve deformable mirror technology to meet the above
      specifications or
      \item improve manufacturing of each optical component in the optical train
      (i.e., super-polishing, super-homogene\-ity, others).
      \end{itemize}

\emph{Multi-object and high-resolution spectrographs:} A requirement
of accuracy in positioning could be $<$1$\mu$m  and will require a
metrology system incorporated in the instrument.
      \begin{itemize}
      \item {\sl Active mirrors} could be required inside the instrument to
      compensate for the variable coma and astigmatism when changing objects
      in the field.  It will also require sensors (interferometer or other)
      to operate in close loop and being able to measure the aberrations with
      accuracies of a few nanometers.

      \item {\sl Adaptive mirrors} could be required inside the instrument
      to compensate for the residual atmospheric turbulence left over large
      field of views.

      \item {\sl Large gratings and filters} in instruments such as WFMOS,
      large VPHs are needed: these are dispersing light by Bragg diffraction from
      periodic modulation of refractive index in a thin layer of processed
      dichromated gelatin (see Sect.~\ref{S4}). The available sizes today are $<$600\-$\times$850 mm.
      Large echelles with diameter $>$1 m could be needed  to achieve very high
      spectral resolution. Large immersion gratings made in Si, Ge,
      ZnSe, a.o.
      will be needed in IR instruments. Large filters will be needed,
      ranging from the visible for LSST, to the near and mid-IR (e.g. MOMSI),
      and in the sub-mm (SCELT).
      \end{itemize}

\emph{Large field correctors and atmospheric dispersion correctors
(ADCs):} In large survey telescopes and in ELTs, large lenses of
$>$1m are needed to compensate for astigmatism and coma across the
field of view. Large ADCs (linear or rotational) are needed to
compensate for chromatic aberrations when changing the zenith
angles. The optical materials of $>$1-m size could be FK5, LLF1,
BK7, fused Silica, ZnS or ZnSE. More candidates are interesting to
explore in detailed ADC designs for specific instruments.

\subsection{CODEX: The Cosmic Deceleration Experiment for the OWL telescope}

\subsubsection{The optics of CODEX}

\begin{figure}
\resizebox{\hsize}{!} {\includegraphics[]{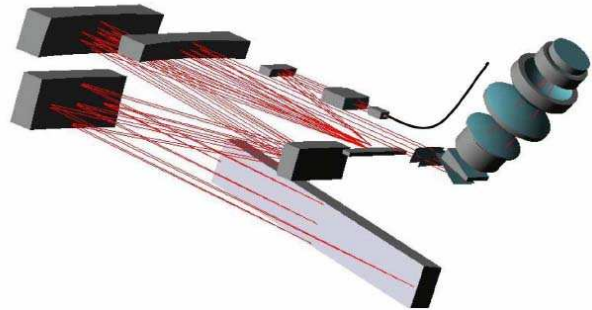}}
\caption{Optical layout of CODEX. The R4 echelle grating is a 4x1
mosaic of total length 1.7m. See text.} \label{fig:delabre1}
\end{figure}

An example of an extremely high stability instrument proposed for an
ELT is CODEX. CODEX is a cluster of $\sim$5 high-resolution
cross-dispersed echelle fiber-fed spectrographs for OWL, working in the
visible spectral range. The instrument operates in seeing-limited
mode. It covers the spectral band between 450 to 680nm at a spectral
resolution in excess of 150,000 (Pasquini et al. 2005). Its main
characteristics are summarized in Table~\ref{tab:delabre1}.
Identical spectrographs are mounted inside a vacuum vessel, hosted
in a temperature controlled room for maximum stability. Several new
concepts have been adopted to ensure high resolution within a
reasonable size of each spectrograph: white-pupil, pupil
anamorphysm, pupil slicing, VPHG cross-disperser. Each spectrograph
is equipped with a R4 1x4 mosaic echelle grating, only twice the
size of those of UVES, but with a 8k$\times$8k detector.
Figure~\ref{fig:delabre1} shows the optical layout of one
spectrograph unit.

\begin{table}[h]
\caption{Main characteristics of CODEX.} \label{tab:delabre1} \scriptsize
\begin{tabular}{ll}\hline
Characteristics & Required value \\
\hline
Sky Aperture      & 1 arcsec for 60m telescope\\
Location          & nested thermally stabilized environment \\
DQE               & 14\% including injection losses \\
Number of unit spectr. & 5 \\
Unit spectr. dimension & diam. 2.4 x 4 m (vacuum vessel) \\
Spectral resolution & 150\,000 \\
Wavelength coverage & 446--671 in 35 orders \\
Spectrograph layout & white pupil \\
Echelle           & 41.6 gr/mm, R4, 170x20 cm, 4x1 mosaic \\
Cross disperser   & VPHG 1500 gr/mm, off-Littrow \\
Camera            & Dioptric F/2.3, image quality 30 $\mu$m \\
Detector          & CCD mosaic 8Kx8K, 15 $\mu$m pixels \\
Noise performance & photon shot noise limited at V=16.5 in 10 min \\
Sampling          & 4 pixels per FWHM \\
\hline
\end{tabular}
\normalsize
\end{table}

\noindent We adopted these design constraints:
\begin{itemize}
      \item No beam larger than 0.5m (need of pupil slicing),
      \item no two dimensional mosaic of grating (need of anamorphic
      beam),
      \item no lenses bigger than 350mm (need of VPH grating).
\end{itemize}

\noindent The instrument will include the following subsystems:
\begin{itemize}
      \item an image slicer,
      \item an anamorphic collimator (ratio 1/4),
      \item a pupil slicer (2 slices),
      \item an echelle spectrograph operating at magnification 1,
      \item a reimaging system including a VPH cross disperser.
\end{itemize}

\begin{figure}
\resizebox{\hsize}{!} {\includegraphics[]{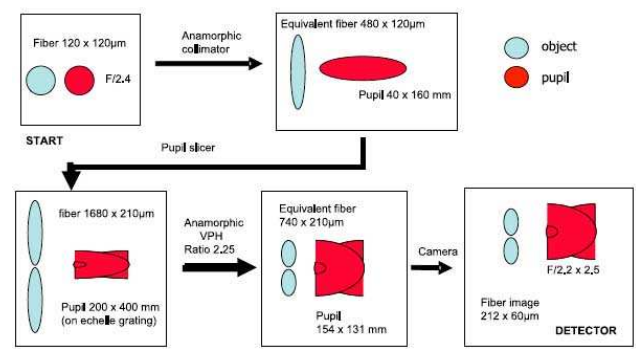}} \caption{Optical
scheme of CODEX.} \label{fig:delabre2}
\end{figure}

The functional diagram of the optical layout is shown in
Fig.~\ref{fig:delabre2}. A 120$\mu$m diameter fiber carries light
from the telescope focus to the spectrograph entrance slit. Short
focal ratios reduce fiber losses due to focal ratio degradation. An
anamorphic collimator with a ratio 1/4 creates an elliptical pupil,
40x160mm in size, onto two tilted off-axis parabolas acting as pupil
slicer. The equivalent fiber image is elongated at the inverse ratio
4:1, i.e. is seen as a 480x120$\mu$m fiber. The pupil slicer mirrors
overlap the two halves of the pupil, reducing its equivalent size to
40x80mm, and create two adjacent fiber images at the F/4.2 entrance
slit of the spectrograph aligned along the height of the slit. This
equivalent slit image is 1680x210$\mu$m. The spectrograph collimator
is a three mirrors anastigmat operating in double pass with an
anamorphic pupil of 400x200mm, where the echelle grating is placed.
This R4 echelle has an overall size of 200x1600mm, i.e., twice that
of the UVES grating. After dispersion, light is focused again near
the entrance slit of the spectrograph, where a pupil relay mirror
sends light toward a transfer collimator of the Maksutov type that
creates a smaller light beam. To reduce the size of the slit images
a volume phase holographic grating (VPHG) acts in a 2.25 anamorphic
magnification, giving an equivalent slit image of 740x210$\mu$m. At
the same time the pupil is compressed at the inverse ratio, giving a
154x131mm pupil. Then the F/2.2 (on-axis) camera reimages the slit
image onto the detector with a final projected size of about
212x60$\mu$m.

The technique of pupil slicing allows the reduction of the grating
area by a factor two (200$\times$1600 instead of 200$\times$\-3200 mm) for the same
resolving power. The price to pay for is factor two for the detector
area. To recover such a spread, a VPHG cross-disperser is used to
enlarge the beam (x2.25) in the cross-dispersion plane, reducing the
detector area by the same factor.

Overall efficiency peaks at 38\% with a 28\% average, from the fiber
exit to the detector focal plane. This estimate is based on the
following assumptions:
\begin{itemize}
\item 99.5\% efficiency  air/glass interface (20x)
\item 98.5\% per reflection (10x)
\item 75\% average for VPH
\item 65\% average for echelle.
\end{itemize}

\subsubsection{Key system components for CODEX}

CODEX will benefit from further development of existing technologies and the
exploitation of new ideas. Here we focus on those components that are crucial
for a successful development of CODEX.

\emph{Light pipe/scrambler:} The illumination of the entrance slit
must be as stable as possible, in order to reduce calibration errors
due to image motions within the slit aperture. This technique has
been successful used in HARPS reaching a very high precision in
radial velocity measurements $<$1 m/s (Avila et al. 2004).

Due to ELT image plane scale, such a scrambler should have a
1x0.12mm cross-section, made in fused Silica. In order to pack
efficiently many fibers together, a light pipe matched to the exit
aperture of some (5-8) fibers is foreseen.

\emph{Echelle grating:} The R4 echelle grating for CODEX will
consist of a 4x1 mosaic of replicas onto the same substrate, with an
overall dimension of 170x22cm. In principle it is a
``straighforward'' extension from the 2x1 mosaic of UVES. However
some aspects need to be further investigated, such as its
dimensional stability (earthquakes, wavefront).

\emph{Beam expander/VPHG:} A key component will be the
cross-disperser, which is also used as a beam expander. We employ a
VPH grating used off-Littrow acting both as crossdisperser and beam
expander. With a useful area of 150x150 mm and a groove density of
1500 gr/mm, such elements do not seem too challenging. Efficiency
and super-blazing properties need to be studied in detail.

\emph{Calibration system:} Metrology labs were recently
revolutionized by the introduction of femtosecond-pulsed,
self-referenced lasers driven by atomic clock standards (Hall \& H\"ansch 2005).
Result is a reproducible, stable ``comb'' of
evenly spaced lines who's frequencies are known a priori to better
than 1 in 10$^{15}$. Advantages are: (a) its absolute calibration,
giving a long term frequency stability required by CODEX; (b) evenly
spaced and highly precise frequencies allowing a mapping of
distortions, drifts and intra-pixel sensitivity variations of CCD;
(c) a naturally fibre feed system.

Some properties need to be developed:
\begin{itemize}
\item Line-spacing currently limited to 1GHz by laser size and energy
considerations. We need $\sim$10--15GHz. New technology needed.
Development project underway with Max-Planck-Institute of Quantum
Optics.

\item Transmission of non-linear optical fibre questionable as  low as $<$400nm.
Existing technology can probably be extended.

\item Moving from the lab to observatory-type environment: "industrial quality"
comb.
\end{itemize}

\emph{Detector Array:} A 8k$\times$8k CCD area will be needed to
correctly sample the echellogram. This corresponds to a
120\-$\times$120mm$^2$ area for a 15$\mu$m pixel size. Each slit image will
be 60$\mu$m$\times$1 mm wide. Three solutions can be envisaged: a
mosaic of four 4k$\times$4k chips, a mosaic of eight 4k$\times$2k
chips, or go to a new generation of very large, wafer-sized
monolithic CCDs, e.g. currently developed for the Potsdam Echelle
Polarimetric and Spectroscopic Instrument (PEPSI) at the
2$\times$8.4m Large Binocular Telescope (LBT). Challenges are posed
by flatness requirements, dimensional and thermal stability ($<$1
mK) and low read noise if light pipe is not feasible.

\subsection{Large lenses and their production}

ELTs will need large optical components to be used in transmission
like lenses, prisms and filters for imaging optics, atmospheric
dispersion correctors and wavelength range filtering. The extreme
dimensions of these components pose challenges for the production,
specification and inspection of the glass blanks.

Between the last mirror in a telescope - may it be a secondary,
tertiary or higher - and the detector, transmitting optics is needed
with large optical elements for beam shaping, atmospheric dispersion
correctors and wave band filters. Even though large transmitting
lenses and prisms have been produced in the past (Johnson 2002) this
cannot be taken for granted. The optical elements will be even
larger and the specifications will be much more stringent reflecting
the ambitious scientific goals of the new telescope generations
(Hartmann et al.~1996; Hartmann et al.~2002; Doehring et al.~2003;
Jedamzik et al. 2004; Doehring et al.~2005). In order to provide
optical elements with the required sizes and qualities in time for
the completion of the telescopes close cooperation between
astronomers and industry will be necessary about the specific
requirements and their verification. The developments needed for
manufacture and metrology methods have to be addressed. Typical for
large optics are long times for manufacturing and for metrology
development. Therefore agreements between the designers for the
transmitting optics and the manufacturers have to be settled early
enough to take the long cycles into account.

The largest piece of optical glass ever made is a 2.15m diameter
350mm thick ZK7 block with a weight of 3.2 tons. It was manufactured
1951 by Jenaer Glaswerke in Jena, Germany by casting the contents of
three clay pots together into one mould. It is still in use, as the
mirror substrate of the 2 m telescope of the Karl Schwarzschild
Observatory in Tautenburg, Thuringia (Hartmann et al.~2004). The
main goal of this cast was the size. Since it was meant as a mirror
blank, optical homogeneity and striae content were not the first
concern. For large transmitting elements in future telescopes these
properties will be crucial. They have to be optimized in production
and measured to highest possible accuracy. Schott has manufactured
blanks with dimensions larger than 1 m and more than 200 mm
thickness with outstanding quality since the 1970s. The striae
content was very low which could be demonstrated by means of the
very sensitive shadowgraph method. The optical homogeneity was
proven by the so called statistical method. Small samples were cut
from the periphery and compared interferometrically (Reitmayer et
al. 1972). The overall refractive index variations were found to be
below 4\,10$^{-6}$ peak to valley and even better. The best results
have been obtained during an optimized serial production run over
several weeks of only one glass type, BK7. Such favorable conditions
occur with projects that need a large amount of high homogeneity
glass. This will be probably not the case for astronomy telescope
aimed production.

\begin{figure}
\resizebox{\hsize}{!} {\includegraphics[]{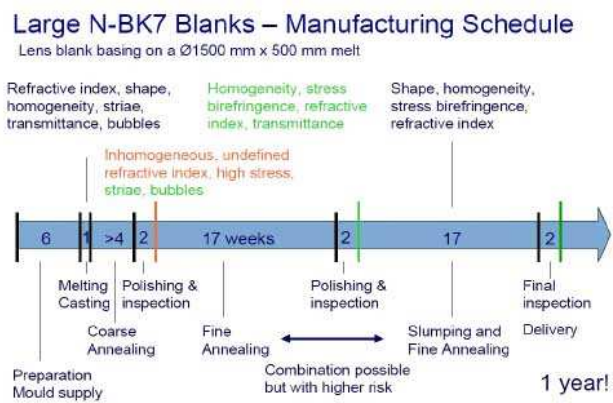}} \caption{Production
sequence of large optical glass blanks. The lower part shows the production
steps, the upper part shows when quality characteristics are defined.}
\label{fig:hartmann1}
\end{figure}

Recently Schott has made an internal assessment of feasible sizes and
maximum volumes. With the large continuous melting
tank providing the best possibility to achieve high optical
homogeneity the maximum dimensions will be 1.5m diameter and 500mm
thickness corresponding to 2.2 tons of N-BK7, the present day
arsenic free version of BK7. Larger diameters with some preforming
of curvatures may be achieved with the slumping method. The glass
will be reheated until it gets soft enough to flow under its own
weight. Standing in a mould with a larger diameter and a curved
bottom, it acquires the new desired shape when sufficient total
volume is available. The total production cycle for a slumped large
blank easily extends to one year. The cooling down in a certain
glass type dependent temperature range, so-called fine annealing,
influences the glass quality decisively, see
Fig.~\ref{fig:hartmann1}.

This production process is the same for all glass types which are
suited for large blank production. This set is restricted mainly to
the classical families: boro-silicate crowns and lead-silicate
flints with some glasses lying close to them in the fluor-crown and
the barium-silicate crown ranges.

This size restriction for optical glass blanks has several reasons:
continuous glass production process is not capable to provide
arbitrarily high glass flows; and glass flow rate must be limited
further to ensure a continuous isotropic flow into the mould to
prevent striae. This leads to long casting times from several hours
up to almost one day. During this time all technical and
environmental parameters have to be kept as stable as possible. The
first stability requirement seems to be self-evident, but it is the
one excluding most glass types from being candidates for large
optics: the glass must remain as a glass. The compositions of many
glasses with exotic optical properties have high tendencies to
crystallize thus preventing production of an integral solid large
piece. The temperature field around the mould must be stable to
prevent shape deformations. The refractive index has to be kept very
stable, since variations in time will be found as spatial variations
i.e. inhomogeneity within the blank. This requires starting with the
production of the glass at least two days earlier to keep away from
the stronger variations typical at the beginning of a melting run.
Temperature controlled coarse annealing preserves the blanks from
breakage due to internal stresses. After a first inspection for
striae and bubbles and inclusions the glass blank will be fine
annealed. This process determines the final refractive index, the
optical homogeneity and the stress birefringence. The refractive
index of a piece of glass is given by its chemical composition only
in the range of 10$^{-3}$. The final values down to the 6$^{th}$ or
7$^{th}$ decimal place will be fixed by the temperature history this
glass piece has undergone in the temperature range from the
so-called transformation temperature to about 150 to 200$^{\circ}$C
downwards. At the transformation temperature, a glass-type specific
value, stress in the glass relaxes within a short time.

During cooling down through this temperature range one must keep the
temperature differences in a large piece of glass as small as possible.
Otherwise this would lead to different refractive indices at different
locations, thus inhomogeneity. Unfortunately glass in general is a poor
temperature conductor, so temperature changes easily lead to high differences
within the volume. In order to keep the differences at minimum, temperature
changes with time have to be kept very slow. Again unfortunately the
differences induced in a volume of a poor thermal conductor are not linear with
its dimensions. They are proportional to the square of the thickness of a
plate. So even if larger blanks could be cast, the annealing times necessary to
achieve high homogeneity would become extremely long, too long to be practical.

\subsection{Glass properties and their measurement}

The main properties essential for the function of a piece of optical glass are:
The refractive index, the Abbe number as a measure of dispersion, the optical
homogeneity, the spectral internal transmittance, bubbles and inclusions,
striae and stress birefringence (Bach et al. 1995; Schott Technical
informations, \textit{www.schott.com}). All are specified in the catalogue and
data sheets for all glass types. However the nominal values and tolerances
cannot just be used or extrapolated to very large pieces of these glass types.
The specially adjusted production processes and process times have influences,
which may interact with each other, changing the specification and which have
to be taken into account therefore.

\subsubsection{Refractive index and dispersion}

\begin{figure}
\resizebox{\hsize}{!} {\includegraphics[]{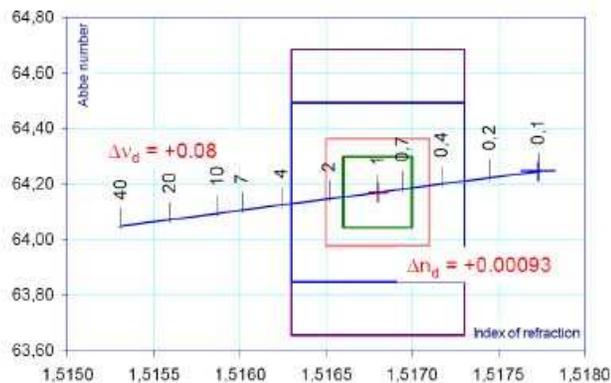}}
\caption{Change of refractive index and Abbe number for large blank
annealing rates of 0.1 K/h for N-BK7 with respect to the catalogue
value (achieved with an annealing rate of 1 K/h). The rectangles are
the catalogue tolerance steps.} \label{fig:hartmann2}
\end{figure}

Very long fine annealing will lead to an increase of the refractive
index (see Fig.~\ref{fig:hartmann2} and table~\ref{tab:hart1}). For
N-BK7 the glass composition for block glass is adjusted for
1$^{\circ}$C per hour fine annealing. Annealing this glass with
0.1$^{\circ}$C/h as necessary for large blanks will increase the
refractive index $n_{\rm d}$ by 0.00093 and the Abbe number $\nu
_{\rm d}$ by 0.08, thus lying out of the tolerance range. The index
d denotes the spectral d-line at 587.6nm. Schott could change the
composition to compensate this effect, but glass produced in this
ways cannot be used for other purposes. So the customer will have to
pay for all preproduction and other ``waste'' glass. If the
instrument optics designer accepts the increased values, the
``waste'' glass, which is not adequate for the large blank but
excellent for smaller pieces may be used for other applications.
Measurement of absolute refractive index and the dispersion is well
understood. The values are determined by taking samples. A precision
test certificate provides the constants of the total dispersion
curve from the near UV to the near IR with an accuracy better than
10$^{-5}$.

\begin{table}[h]
\caption{Change of refractive index and Abbe number for annealing rate ratio
1:10 for some optical glasses.} \label{tab:hart1}
\begin{tabular}{ccccc}\hline
Glass type  & $n_d$     & $\nu_d$   & $\Delta n_{d(1:10)}$ & $\Delta \nu_{d(1:10)}$ \\
\hline
LLF1   & 1.54814 & 45.75 & 0.00040 & 0.02 \\
LLF6   & 1.53172 & 48.76 & 0.00045 & 0.03 \\
F2     & 1.62004 & 36.37 & 0.00043 & 0.04 \\
SF6    & 1.80518 & 25.43 & 0.00058 & 0.05 \\
N-BK7  & 1.51680 & 64.17 & 0.00093 & 0.08 \\
\hline
\end{tabular}
\end{table}

\subsubsection{Optical homogeneity}

The striae inspection of the glass blank after coarse annealing gives a first
assessment of the homogeneity achievable. If there are only very few and faint
striae, there is a high chance that the homogeneity will be excellent after
fine annealing. But this is only a necessary precondition, not a guarantee. The
success of production will be proven only after lengthy fine annealing, half a
year after casting.

The common way to specify homogeneity in a catalogue is to give a
limit for the peak-to-valley (p-v) value. This value may be
misleading to a certain extent. It comprises contributions to the
wave front deviation which are not critical for the function of the
optical element. Especially the defocusing term can easily be
corrected in the optical system. Also astigmatism may be compensated
by the rotation of other astigmatic elements in the system to a
certain extent. This may allow relaxation of the homogeneity
specification, because the terms mentioned normally contribute a
significant amount to the p-v value. Adapted polishing may
compensate other long-range deviations, additionally. So a close
communication between the glass manufacturer and the polisher will
be very helpful.

The measurement of the optical homogeneity is the most challenging
inspection to be made. The Fizeau interferometers presently used for
this purpose have a maximum aperture of 500 to 600mm. For elements
up to an effective diameter of 1.5 m there will be no chance to
measure them with a single large aperture. In order to build such an
interferometer lenses of 1.5 m would be needed -- a bootstrapping
problem. Other influences like the variations in the environmental
conditions like the temperature field around the blank to be
measured and around the total interferometer with its large optical
elements, vibrations and air flow in the interferometer cavity also
limit the maximum possible aperture.

The wave front deviation  $\Delta W$ for plane waves traveling
through an optical transparent material with thickness $t$ and
temperature inhomogeneity $\Delta T$ is calculated according to
Reitmayer \& Schroeder (1974) with Eq.~\ref{for:hart1},
\begin{equation}
   \Delta W = t \left( \alpha (n(\lambda) - 1) + \frac{dn}{dT} \right) \Delta T
\label{for:hart1}
\end{equation}
where $\alpha$ is the coefficient of thermal expansion, $n$ is the
refractive index, and $\frac{dn}{dT}$ is the thermo-optical
coefficient of the glass type. The extreme temperature sensitivity
of thick pieces of optical materials can be seen by the values
calculated with Eq.~\ref{for:hart1} and given in
Table~\ref{tab:hart2}.

\begin{table}[h]
\caption{Wavefront deformation of plane waves caused by temperature
inhomogeneity.} \label{tab:hart2} \footnotesize
\begin{tabular}{lrrrrr}\hline
Material & $\alpha$ & $dn/dT$ & $n$  &  $G$ & $\Delta
         W$ \\
         &  (-30/70$^{\circ}$) & (20/40$^{\circ}$) &   &  &  \\
         & 10$^{-6}$  & 10$^{-6}$  & & 10$^{-6}$  & \\
         & (K$^{-1}$) & (K$^{-1}$) & & (K$^{-1}$) & (nm) \\
\hline
LLF1 & 8.10 & 2.90 & 1.55099 & 7.36 & 147 \\
LLF6 & 7.50 & 3.70 & 1.53431 & 7.71 & 154 \\
F2   & 8.20 & 4.40 & 1.62408 & 9.52 & 190 \\
SF6  & 8.10 & 11.10& 1.81265 & 17.68 & 354 \\
N-BK7& 7.10 & 3.00 & 1.51872 & 6.68 & 134 \\
Vitr. Silica & 0.51 & 10.10 & 1.46008 & 10.33 & 207 \\
Zerodur & 0.02 & 14.80 & 1.54470 & 14.81 & 296 \\
Air  & - & -0.92 & - & -0.92 & -18 \\
\hline
$\alpha$ & \multicolumn{5}{l}{ constant of thermal expansion} \\
$n$ & \multicolumn{5}{l}{constant of thermal expansion} \\
$G$ & \multicolumn{5}{l}{thermo-optical coefficient} \\
$\Delta$W & \multicolumn{5}{l}{wavefront deformation} \\

\end{tabular}
\normalsize
\end{table}

This holds not only for the quality inspection of large blanks but also for
their application as optical elements in the telescope. Temperature changes
transformed by transmitting materials into wave front deformations will be a
significant source for errors.

So there will be no way around stitching sub-apertures. This has been done
already (Sch\"onefeld et al. 2005) but only for much smaller sizes. Especially
for the specification and verification of the homogeneity it is urgent for the
telescope optics designers and the glass manufacturers and the polishers to
agree on the values needed. If new larger interferometers and evaluation
methods have to be developed, one must take into account long development times
(several years) and high costs.

\subsubsection{Internal transmittance}

\begin{figure}
\resizebox{\hsize}{!} {\includegraphics[]{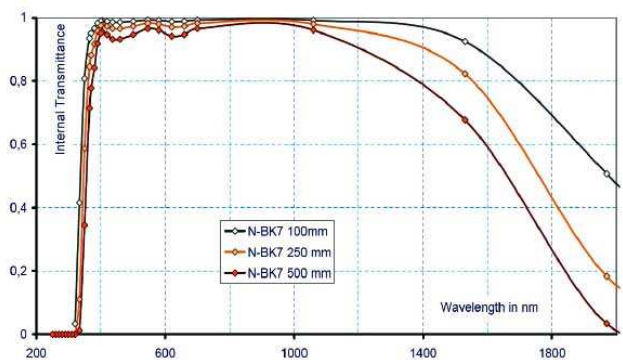}} \caption{Spectral
internal transmittance of N-BK7 for three different thicknesses.}
\label{fig:hart3}
\end{figure}

The internal transmittance in the visible range is normally no big
concern even for large thicknesses (see Fig.~\ref{fig:hart3}).
However travelling into the IR wavelength range the internal
transmittance starts to decrease. Large thicknesses will pronounce
this effect. Measurement using samples is common practice and is
representative for a large blank.

\subsubsection{Bubbles and inclusions}

Candidate glass types for large lenses have typically a very low content of
bubbles and inclusions. But the existence of some scattered bubbles cannot be
excluded totally. Usually one specifies the ratio of the sum of the effective
areas of all inclusions to the area of the optical element. This ratio is a
measure of the stray light produced in the element.

\subsubsection{Striae}

Striae are variations of the refractive index in short-ranges with
typical periods between 0.1mm and 1mm. Their main effect is again
stray light. One possibility to specify striae is to limit the ratio
of the total area covered by striae to the total optical element
area disregarding the wave front distortion p-v value. This was
applied in the case where some few striae with sharply defined areas
existed, when optical glass was made mainly in clay pots. Nowadays
the appearance of striae has changed. They may cover larger parts of
the total element volume and thus area, but are much smaller with
respect to the wave front distortion. Since their wave front
distortion effect is proportional to the light path to a certain
extent, for common optical glass standard striae quality has been
defined to be better than 30nm optical path difference per 50mm path
length. The candidate glass types for large lenses have the
potential of much better quality as is shown by practical
experience. To keep the striae content on a low level will be really
a challenge with large glass blanks. Especially close to the edges
striae can appear. Since part of the optical elements edge zone will
be hidden by mounting assemblies, there may be some room for
relaxation of the striae tolerance in this outer zone.

\subsubsection{Stress birefringence}

Stress birefringence is the other reason for the very long annealing
times. The temperature differences in the blank's volume produced
during the ramp down in the range below the transformation
temperature will result in permanent bulk stresses. Only lowest
annealing rates and smallest possible thickness help in getting best
stress birefringence results. The standard tolerance for small
optical elements will probably not be sufficient. 10 nm/cm
birefringence in block or strip glass is harmless since most
elements made out of such glass have only optical path lengths of
some cm and additionally stress collapses significantly while
cutting blocks to smaller pieces. For large optical glass blanks
birefringence will become significant. There is no cutting to reduce
bulk stress and the large optical path lengths sum up birefringence
to significant amounts. An element with a thickness of 300mm and 10
nm/cm specific birefringence would end up with 300-nm birefringence
in total. The optical design would have to take into account wave
front retardations of 300nm between orthogonal polarized light rays
at maximum varying locally over the elements area. For a
well-annealed piece of glass bulk stress is highest at the edge,
where it is compressive. It decreases towards the centre, crosses
zero and reaches a local maximum at the centre, where it becomes
tensile. The effective birefringence results from the interaction of
the light with its different polarization directions and the bulk
stress tensor field, which makes things very complicated. So the
best will be to reduce stress birefringence in total to the best
possible value, this is limited by practically possible annealing
times.

\subsection{Large filters and their production}

In principle there are two ways to produce large filters, by
interference coatings on white or colored substrate glass sheets or
by using colored glasses. Facilities for the production of large
coated filters have already been developed (Geyl 2004). These
filters can be tailored for a very well defined spectral
transmittance with sharp edges between the transmitting and blocking
ranges. However this holds only for light with normal incidence.
With increasing angles of incidence  $\alpha$ the transmitting
ranges shift towards shorter wavelengths significantly,
approximately proportional to $\sin^2\alpha$. Colored glass filters
do not have edges as steep as is possible with coated filters
especially at the long wavelength cut-off, but they are much less
sensitive to the incidence angle.

\begin{figure*}
\resizebox{\hsize}{!} {\includegraphics[]{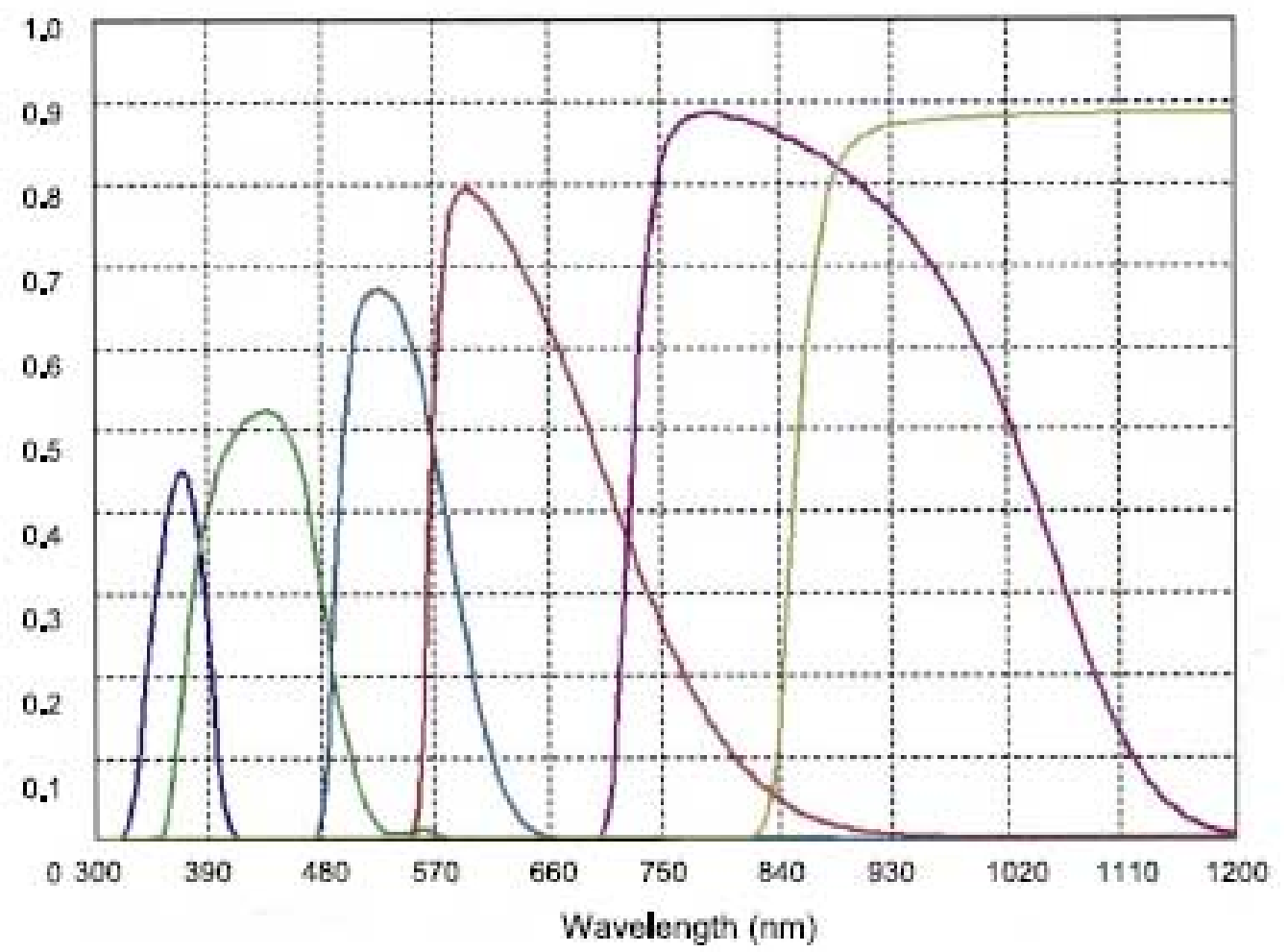} \includegraphics[]{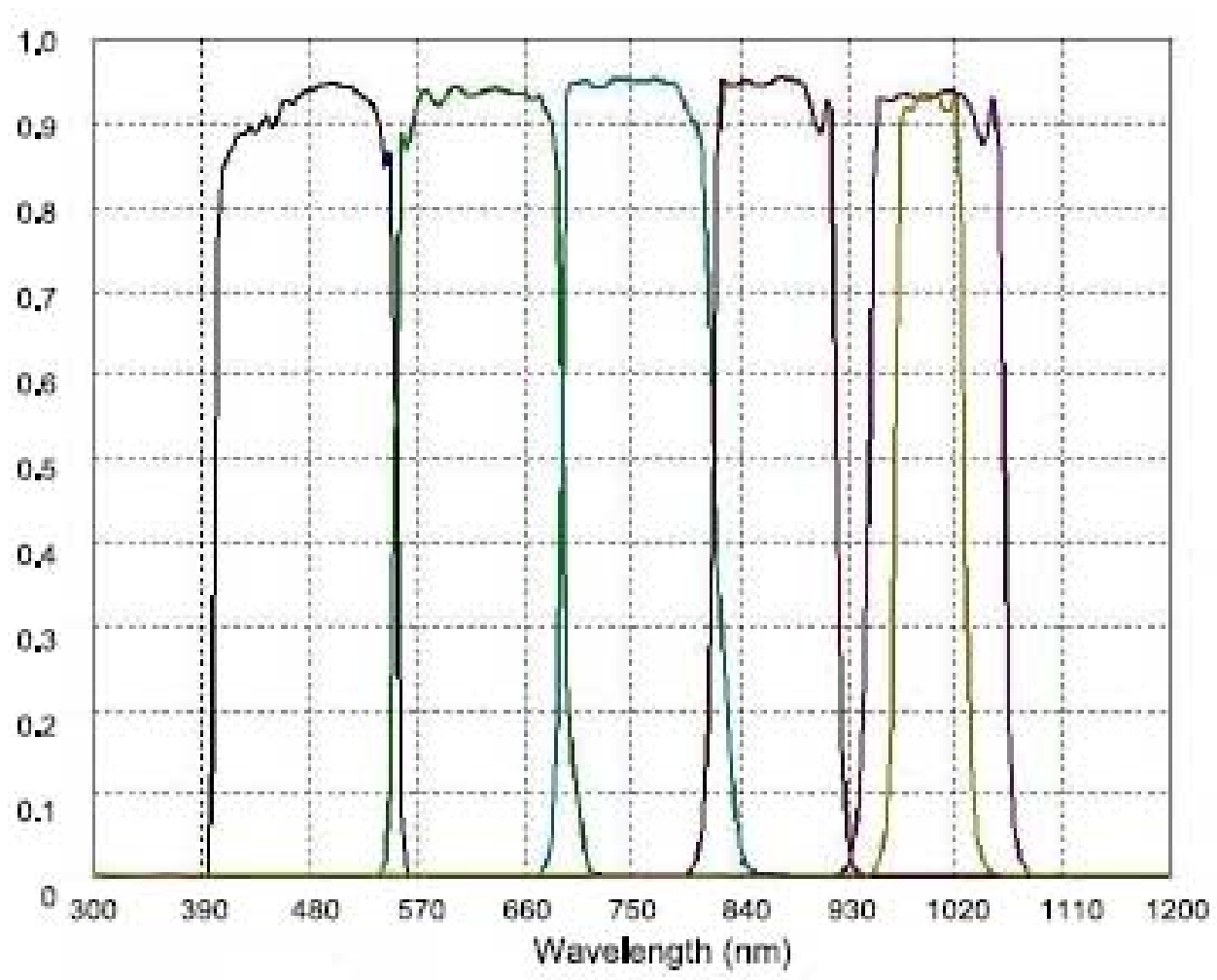}}
\caption{Spectral transmittance for a Sloan filter set from of
colored glass (left) and from coated filters (right).}
\label{fig:hart4}
\end{figure*}
\begin{figure*}
\resizebox{\hsize}{!} {\includegraphics[]{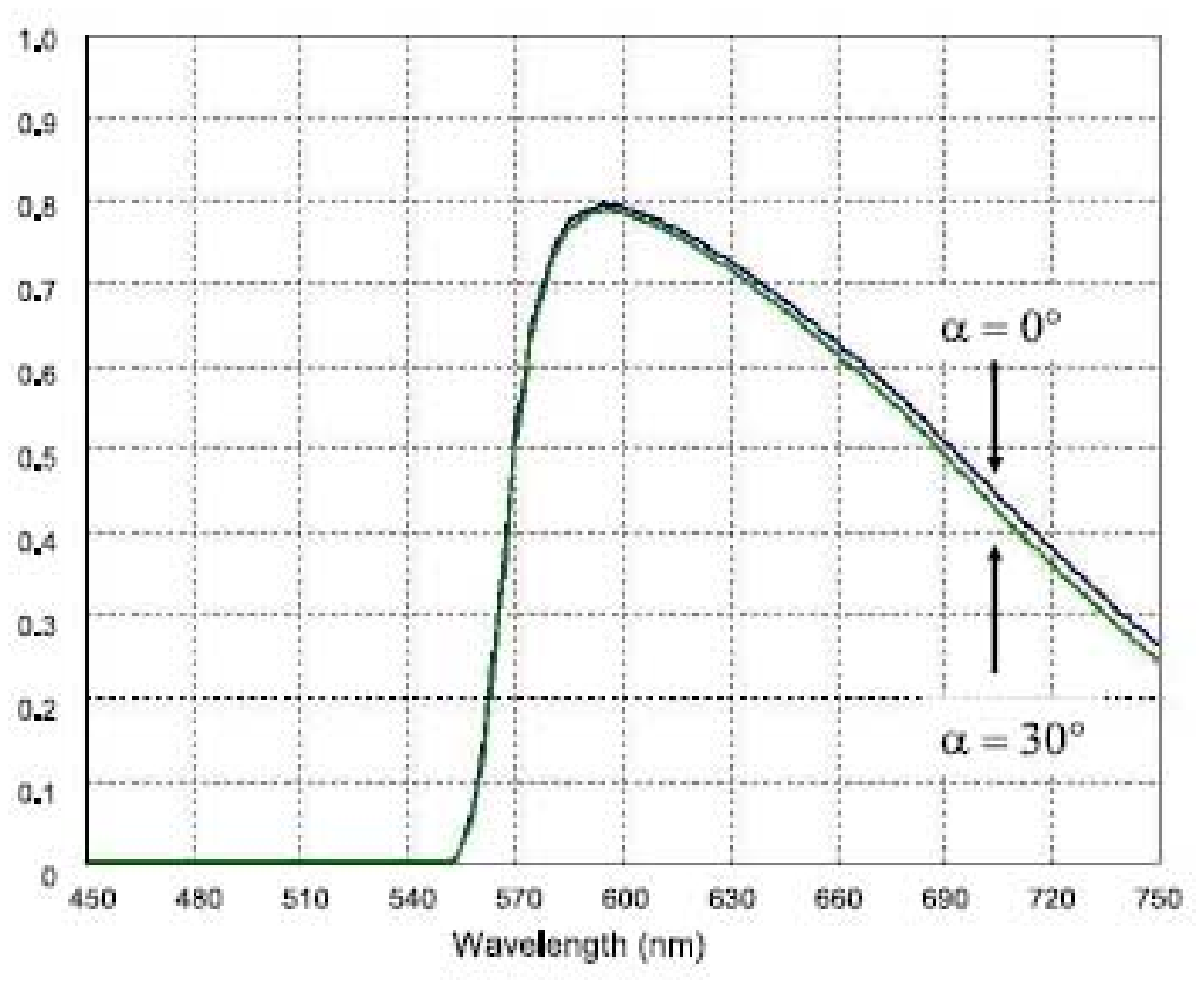}\includegraphics[]{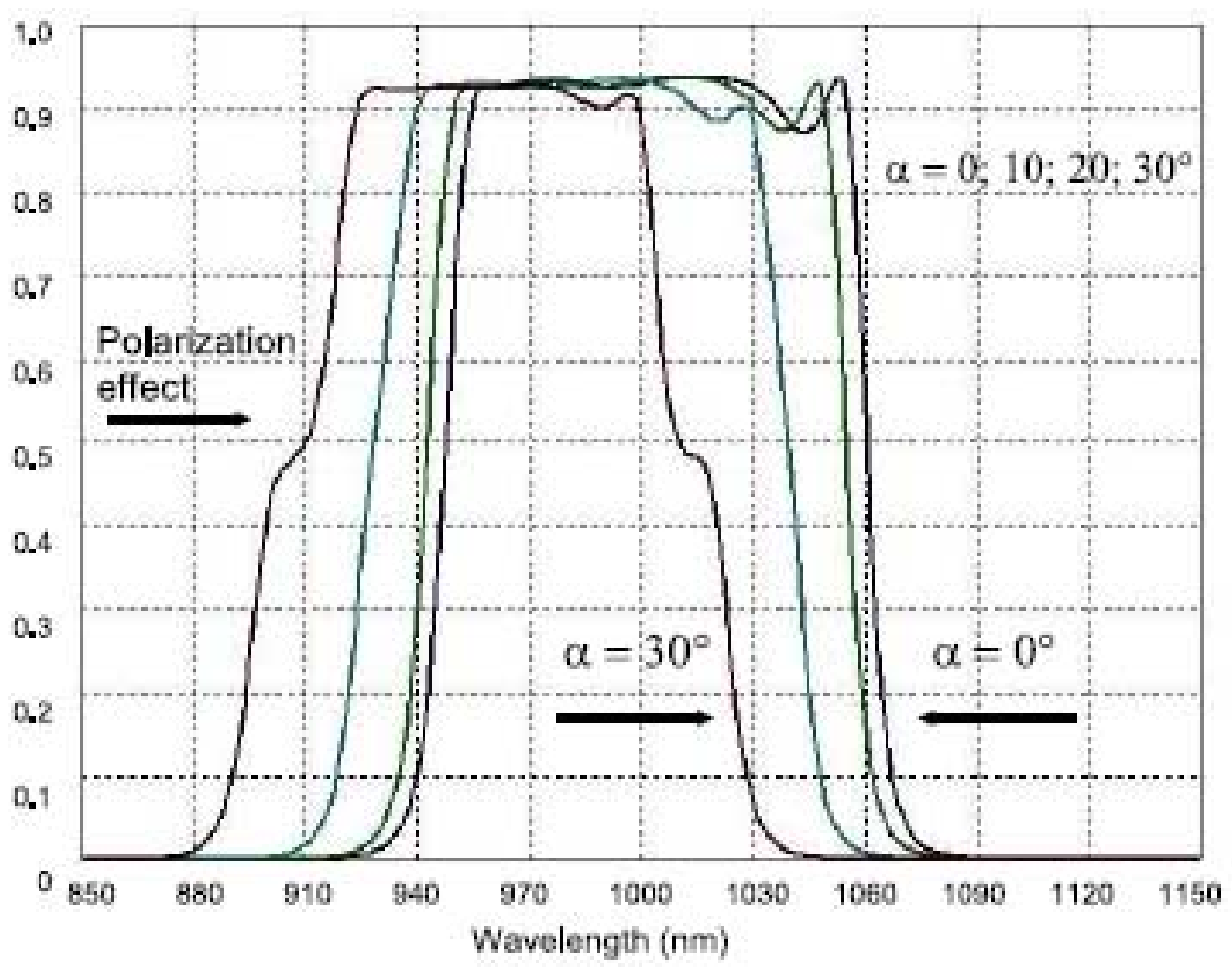}}
\caption{Dependence of the spectral transmittance on the angle of
light incidence for a band filter of colored glass (left) and from
coated filters (right).} \label{fig:hart5}
\end{figure*}

Instrument designers need sets of different glasses (11 types for the Sloan
filter set), which are restricted in size in different ways. Some glass types
are produced as large sheets. Others with higher crystallization tendencies
would need development to achieve larger dimensions. Since these developments
will not just be simple extrapolations, their results are at present
undetermined. Furthermore there would be a very bad ratio between the efforts
and the possible revenues, so that a commercial company would not do without
considerable financial inducement.

Presently available or possible sizes are shown in
Table~\ref{tab:hart3}. All colored glass types are produced only
once in a while since with one production run one obtains the need
for one year or even more.  The large 900x900mm$^2$ sheets of the
GG, OG and RG types are produced only once a year. If there is no
special requirement, they will be cut down to small pieces. The vast
majority is sold as 50x50mm$^2$ plates. Hence there are time windows
where special requirements can be taken into account.

\begin{table}[h]
\caption{Coloured glass types for the Sloan filter set and their possible
dimensions.} \label{tab:hart3} \scriptsize
\begin{tabular}{ccccc}\hline
Filter & Type of  & Current sizes & Max. sizes & Notes \\
glass  & coloring & produced      & current capab. & Larger sizes \\
       &          & (mm x mm)     & (mm x mm)      &  \\
\hline
UG1    & ionic. & 220x220 & 240x240 & R\&D \\
BG12   & ionic. & 220x220 & 240x240 & discontinued \\
BG18   & ionic. & 240x240 & 360x240 & R\&D \\
BG39   & ionic. & 240x240 & 360x240 & R\&D \\
KG3    & ionic. & 625x185 & 800x280 & R\&D \\
RG9    & ionic./coll. & 360x360 & 900x900 & available \\
GG385  & coll.  & 360x360 & 900x900 & available \\
GG495  & coll.  & 360x360 & 900x900 & available \\
OG570  & coll.  & 360x360 & 900x900 & available \\
RG850  & coll.  & 360x360 & 900x900 & available \\
N-WG280 & base glass & 200x200 & 280x280 & R\&D \\
\hline
\end{tabular}
\normalsize
\end{table}

Because in the past there were no requirements for glass filters
beyond 300mm in diameter only little is known about the internal
quality of pieces of these sizes. Optical homogeneity and refractive
index homogeneity probably will not be a significant problem because
of the small thickness of the filters. The optical transmittance
homogeneity is not known. First tests will be made soon for the GG,
OG and RG types presently being produced. Stress birefringence also
will be no problem due to the small thickness. But there may be
problems with bubbles and striae. Zones with reduced quality cannot
be cut out and thrown away as in the production of small filters.
From the production point of view mosaics from hexagonal filter
elements would be much easier, much cheaper and much better to be
controlled for their quality. It will be worthwhile to recheck the
necessity of large monolithic filters in any case.

\subsection{Free-form optics}

The segments for extremely large telescopes present unprecedented challenges in
industrial-scale manufacturing, surface-spe\-cification, and metrology.
Quality-assurance over a long produc\-tion-cycle is also an issue. The adoption
of a segmented spherical primary mirror goes some way to alleviate the
problems, but at the expense of a complex optical system and impaired
stray-light and infrared-emissivity – both of which are detrimental to key
science-goals such as detection of extra-solar terrestrial planets. Simple
two-mirror telescope designs are clearly superior, but at the expense of
off-axis aspheric segments.

Emerging new technologies can be brought to bear on this challenge.
The United Kingdom is investing in a new ``National Facility in
Ultra-precision Surfaces'' which is including a 1m pilot segment
production plant for aspheric process-development. Moreover, an
automated polishing of free-form surfaces has been set up, which
could have a major impact on the optical design of ELT
instrumentation.

ELTs will benefit from the ability to produce $\approx$1m-size
segments, both for the telescope mirrors (primary, secondary,
tertiary) and related instrumentation. Such large optical machines
put stringent requirements on:
\begin{itemize}
\item Delivered point spread function (PSF),
\item surface power spectral density (PSD), and
\item edge control (segments).
\end{itemize}

A large project started under the UK Research Council's Basic
Technology Initiative, called ``Ultra--Precision Surfaces'' and led
by University College London and Cranfield University. Its first
objective is to establish a national facility where capabilities for
\emph{complex}\footnote{i.e., non-rotationally symmetrical and
completely \emph{free-form} surfaces} surfaces will be developed,
with a substantial improvement in ratio of cost/surface-accuracy.

\begin{figure}
\resizebox{\hsize}{!} {\includegraphics[]{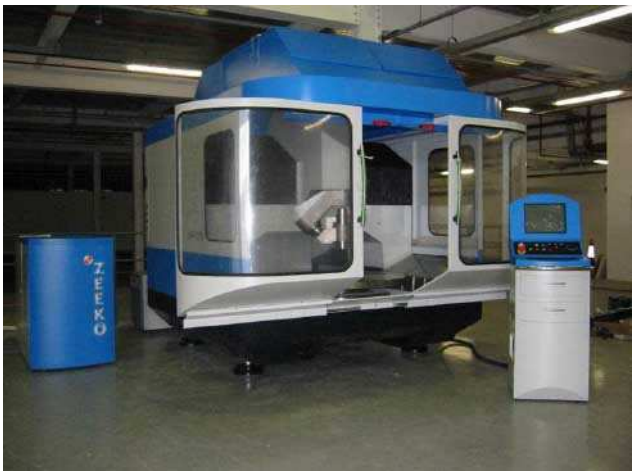}} \caption{Zeeko 1200mm
CNC Classic/Fluid-jet polishing machine.} \label{fig:walker1}
\end{figure}

The technical approach will be optimally to combine several novel
processes and manage in a deterministic way residual surface
structures to optimize performances. Within these new technologies
we have: computer numerical control (CNC) membrane local-polishing,
CNC fluid-jet polishing, CNC ``grolishing'' (a hybrid
grinding-polishing process), ultra-precision grinding
(Fig.~\ref{fig:walker1}), reactive atomic plasma technology, and
surface metrology.

\begin{figure}
\resizebox{\hsize}{!} {\includegraphics[]{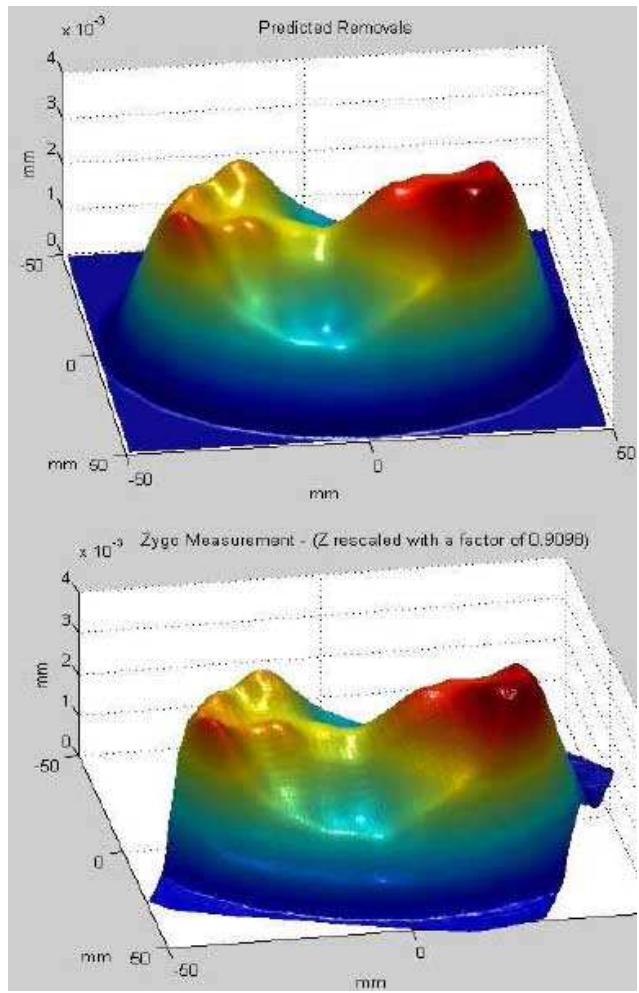}}
\caption{Free--form polishing results: predited (top) and measured
(bottom).} \label{fig:walker2}
\end{figure}

One of the most interesting technologies is the \emph{free-form}
polishing. Predicted and measured results are in agreement as shown
in Fig.~\ref{fig:walker2}. This technique has a potential impact on
instrument design. A complex optical system can be simplified and/or
enhanced by giving the designer additional mathematical degrees of
freedom. Moreover, fewer optical surfaces with superior image
quality, less stray light, and lower infrared emissivity will
improve optical performances (e.g., off-axis mirror systems).
Finally, it can be used to correct system aberrations on a surface
near a pupil image.

\section{Metrology for optics}

This science of precision measuring of physical parameters is the
most important part of the ELT project.  High precision and
efficient metrology is required in the manufacturing of thousands of
large off-axis segmented mirrors. The testing is the bottleneck of
the process today not the polishing. Manufacturing large convex
secondary mirrors ($>$5m diameter) is difficult and designers are
proposing Gregorian telescope increasing the size of the telescope
structure because they think that the metrology and testing cannot
achieve the accuracy required. We need to invest in brainpower and
budget to find ways to overcome these difficulties. In
instrumentation there is a huge challenge because the metrology
systems (interferometry, profilometry) would be an integrated part
of the instrument, have to operate in close loop, be in cryogenic
environments, and requiring $\mu$m or even nm levels of accuracy.

\subsection{Testing aspherical surfaces}

The importance and the utility of testing aspheres goes beyond the important
task of quality assurance and optical component certification. Indeed the
popular statement ``you can make what you can measure'' is particularly true in
asphere manufacturing. During the fabrication process frequent surface
measurements are needed  to determine what is wrong and what can be done to
make the surface better especially when it is not possible to maintain under
strict control all the parameters affecting the fabrication process.

There are two different measurement methods to test aspherical surfaces:
profilometry and interferometry. Even if large profilometers exist,
interferometers are often preferred, allowing us to measure the whole 3D shape
of a surface. However interferometry with aspheres poses many ``new'' problems:
only part of the beam coming back from the surface will enter the
interferometer and will be measured; fringes will be densely packed within the
interferogram, asking for very high spatial resolution of the instrument. If
fringes cannot be resolved, it is safer to look for optical configurations to
``null'' the fringes, inserting some optical components or change the test set
up in order to finally resolve fringes.

There are no general recipes to find null lens configurations. Simulations can
be helpful to understand if the testing configuration is working properly. In
some configurations, test rays are not normal to the surface, then the
interferogram will not be directly related to surface errors, but will have to
be decoded before to be given to the optical workshop for further polishing.
Off-axis conical surfaces can be even more difficult to characterize in term of
its radius of curvature and conic constant: a good relative error for the
latter will often be larger than few thousandths.

\begin{figure}
\resizebox{\hsize}{!} {\includegraphics[]{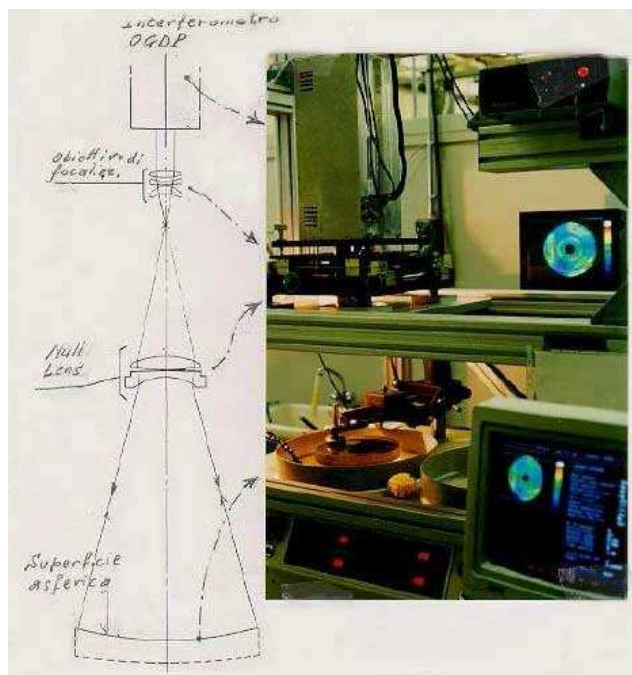}} \caption{Null lens test
setup to test a quartz asphere during final polishing.} \label{fig:melozzi1}
\end{figure}

A well known class of conventional null lens are those generated by
the ``Offner'' configuration that can be a good starting point.
Refractive null lenses are easier to build by the optical workshop
due the availability of tools. A typical application is the
continuous measurement of surface profile during final polishing
(Fig.~\ref{fig:melozzi1}). Convex surfaces are more and more
difficult to measure. Unless for conical surfaces (hyperboloid,
ellipsoid), polynomial convex surfaces will require ``near'' null
lens tests, but particular care must be applied to reduce residual
errors coming from higher order aberrations.

An issue is how to ``test'' the null lens: sometimes, when you are testing
several identical aspheres, a constant error in the interference fringe pattern
can probably be attributed to the null lens itself. When only one surface must
be done, computer generated holograms (CGH)can help in simulating the wavefront
from the asphere. This technique enables obtaining a surface accuracy
(peak-to-valley) of the order of $\lambda$/10, but positioning errors of the
CGH must be taken into account.

There are also some tricks, like sub-Nyquist interferometry. This
requires less than two pixels per fringe, assuming that first and
second surface derivative is continuous. It requires a custom CCD or
a pinhole mask, with strong alignment problems. Other solutions are:
\begin{itemize}
\item Infrared interferometry for testing pre-shaped (rough)
surfaces,
\item two wavelength holography (633-nm and 543-nm lasers), and
\item high density arrays.
\end{itemize}

However, a ``universal'' method to test aspheres does not exist and the
engineer should evaluate case-by-case the most appropriate approach as there
are several possible setups for testing aspherical surfaces and wavefronts.


\section{New materials and innovative processes}\label{S4}

Traditional astronomical instrumentation has been mostly designed
and built in the last 50 years using metal and glass. Detailed
Finite Element Models (FEM), light-weighting techniques, clever
optical designs and active controls, allowed scaling up instruments
from the 2-m class telescopes to the 10-m class telescope without
major modifications in the materials used.

A further scale-up to ELT pupil sizes for seeing-limited instruments
is not straightforward first of all because it would turn
instruments into very heavy systems difficult to control and to move
around with the telescope. Furthermore the requirement on optical
component will be moving from ``perfect single piece'' to
``acceptable series production'' this because many of the
instruments currently sketched for ELTs are made of a number of
identical subsystems.

Materials such as Polymeric compounds, composites, light metals,
functional alloys a.o. will likely be progressively introduced into
astronomical instrumentation design. Accordingly innovative
processes such as ion beam figuring, holography, etc. will as well
progressively replace traditional manufacturing techniques.

This section, reporting the discussion of this session of the workshop, intends
to assess the status of the art in the above fields and serve as a baseline for
future joint institute-industry developments.

\subsection{Production of large-size volume phase holographic
gratings}\label{VPH_blanche}

Efficient dispersing elements are of great interest in astronomical
instrumentation; moreover the possibility of making large size
gratings fits the needs of new instrumentations for current large
telescopes and the next generation ELTs. Volume phase holographic
gratings (VPHGs) are good candidate for these aims (Barden et al.
1998, Barden et al. 2000, Blanche et al. 2004).

VPHGs are dispersive elements as are conventional surface relief
gratings but, instead of having dips and bumps, the diffraction is
achieved by refractive index modulation of the bulk material (Smith
1977; Schankoff 1968; Chang 1979; Hariharan 1984). So, light coming
from the instrument passes through the grating and the dispersed
spectrum is behind the grating.

Due to their intrinsic structure, volume phase gratings possess
unique properties (Kogelnik 1969; Barden et al. 1998):
\begin{itemize}
  \item They have a theoretical efficiency of 100\%.
  \item The undiffracted part of the spectrum is not perturbed by the grating
  and can   be reused for other applications. In this configuration, the grating
  acts like a filter.
  \item Glued between two prisms, the diffracted spectrum is not deviated
  which allows  to have a straight pass spectrometer: the Littrow configuration.
  These   prisms/grating elements are named ``grisms''.
  \item The superblaze property: Changing the incidence angle of the light
  arriving on the VPHG shifts the blaze wavelength. Envelope of all the blaze
  curves for   different angle is called ``superblaze''. This property broaden
  the useful bandwidth.
  \item Gratings are recorded by holographic setup. This allows to record
  large size   in one single laser shot instead to rule each line one by one.
  Thus, each grating is a master.
  \item Sturdiness: the active layer (few microns of dichromated gelatine)
  is encapsulated between two blanks. VPHG elements can so be handled and
  cleaned as regular piece of glass.
\end{itemize}
These advantages make VPHGs the best choice as spectrometer
dispersive element and this is why they are more and more used by
astronomers when it comes to design new instruments (Arns et al.
1999; Monnet et al. 2002; Blanche et al. 2002; Andersen et al.
2004).

For tens of years, the Centre Spatial de Li\`ege had the know-how to
produce this particular type of grating. In year 2000, a new
facility was set up which allows both research and manufacturing of
large VPHGs (Habraken et al. 2001, 2002; Blanche et al. 2002). In
2005, in order to distribute these elements, the ATHOL company was
founded (acronym for ``Advanced Techniques in HOLography'' for
Optics).

Manufacturing VPHGs at ATHOL requires three steps:
\begin{itemize}
 \item Coating the active layer,
 \item recording the grating with holographic setup, and
 \item processing the film to develop the index modulation.
 \end{itemize}
Of course, it is also required to check and certify all the
parameters of the gratings after production, such as line density,
efficiency, blaze wavelength, bandwidth, and wavefront errors.

\subsubsection{Monolithic gratings}

ATHOL coats own dichromated gelatine from 2$\mu$m up to 25$\mu$m.
Thickness is certified by grooved spectra measurement. The blank
size the whole system can accommodate is about 500~mm a side. The
clear aperture of the recording setup is 380~mm of diameter. The
capabilities are VPHGs with fringe frequency from 300~lines/mm up to
more than 3500~lines/mm. These limits can be overcome for gratings
of smaller size but the horizontal dimension can also be enlarged
when the beam is made oval during the casting on to the recording
plane.

\begin{figure}
 \resizebox{\hsize}{!}
 {\includegraphics[]{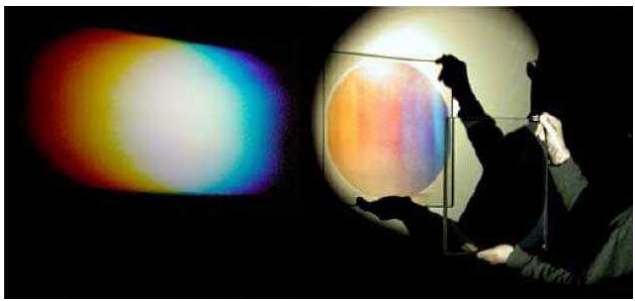}}
 \caption{380~mm diameter monolithic VPHG from ATHOL.}
 \label{fig:GolemVPH}
\end{figure}

The largest monolithic VPHG made so far at ATHOL was a 380-mm
diameter grating for the Osservatorio Astronomico di Brera in Italy.
Figure~\ref{fig:GolemVPH} shows this grating diffracting a beam from
a halogen light. The first order spectrum is diffracted on the right
and the colored circle is the zero order which contains the
complementary colors since nearly 100\% of the light at the blaze
wavelength is diffracted.

\begin{figure}
 \resizebox{\hsize}{!}
   {\includegraphics[]{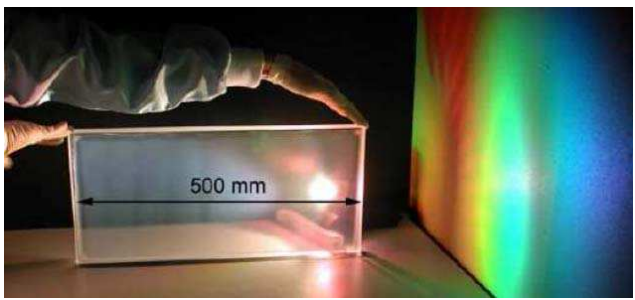}}
 \caption{500~mm wide monolithic VPHG from ATHOL.}
 \label{fig:WIYN}
\end{figure}

Figure~\ref{fig:WIYN} presents a monolithic grating produced for
NOAO where ATHOL took advantage of the enlargement of the recording
beam horizontal axis. Indeed, with a recording angle of $60\degree$,
beams are made to be oval on the substrate and so enhance the size
to 500~mm.

But whatever the size of the facility, it will always be finite and
this restrains the instrument designer's creativity. To overcome
this problem first tests with the the mosaic technique were
performed.

\subsubsection{Mosaics}

The mosaic technique consists of assembling several gratings
recorded and processed independently. Challenges are that mosaic
subelements have to diffract at exactly the same angle and according
to the same blaze and superblaze. That means that gelatine thickness
and index modulation have to be perfectly matched from element to
element.

Moreover, during the encapsulation, elements have also to be
co-aligned precisely in order to avoid any tilt between recorded
fringes. Such an angle will induce diffraction by the elements in
different directions.

\begin{figure}
 \resizebox{\hsize}{!}
  {\includegraphics[]{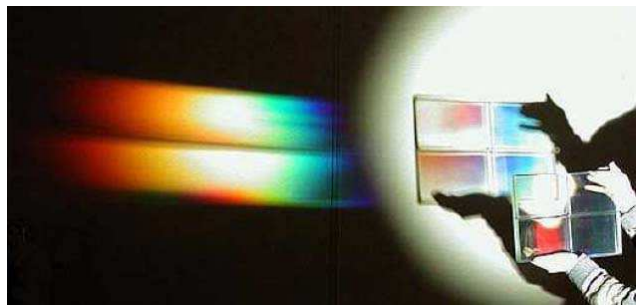}}
 \caption{Four elements mosaic VPHG from ATHOL.}
 \label{fig:mosaic}
\end{figure}

Two mosaics were produced for NOAO, both consisted of four elements
for a total size of 340x240~mm. Figure~\ref{fig:mosaic} shows one of
these mosaics. Note that spectra perfectly overlap, proving the good
alignment of the subgratings. The dark middle line appearing in the
spectrum is due to beam baffling by the blanks beveled edges. Thanks
to the mosaic technique, the size of the setup no longer matters and
the final grating diameter becomes virtually unlimited.

\subsubsection{Cryogenic operation}

VPHGs can be manufactured so that they diffract in the infrared. Up
to now, ATHOL made gratings diffracting up to 2.5 microns. However,
in spite of their infrared effectiveness, their actual use is
restricted to spectrometers running at ambient temperature. This is
due to lack of knowledge of their behavior in cooled IR
spectrometers which have to operate at cryogenic temperature.

To use VPHGs as IR dispersive elements, one has to ensure they will
survive. Experiments were made at ATHOL that placed samples in a
cryogenic vacuum vessel and measured efficiency as well as
diffracted wavefront at ambient temperature as well as down to
liquid nitrogen temperatures.

\emph{Efficiency:} Rather to measure the diffraction efficiency of
the +1 order, the 0 order transmission efficiency is measured. Using
a fiber spectrometer allows to record the blaze curve in one single
measurement and to avoid a light collection problem into the fiber.

Ensuring there is no higher order, +1 and 0 order efficiency are
related by the following relation
\begin{equation}
\eta_{+1} = 1 - \eta_{0} - L
\end{equation}
where the losses $L$ take into account Fresnel reflection,
absorption, and diffusion. Thus, such measurements record an
efficiency dip instead of being an efficiency bump.

\begin{figure}
 \resizebox{\hsize}{!}
  {\includegraphics[]{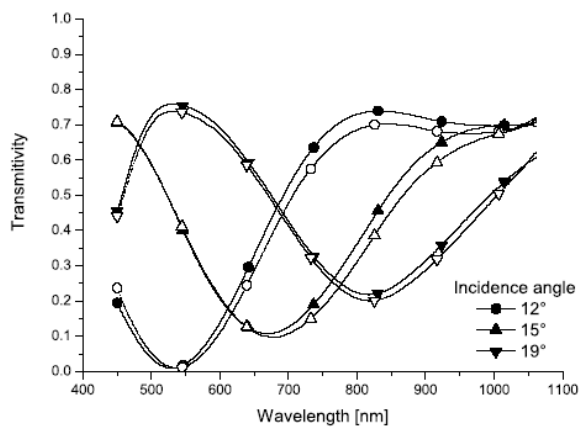}}
 \caption{Transmission efficiency at ambient (plain symbols) and cryogenic
 (hollow symbols) temperature.}
 \label{fig:efficiency}
\end{figure}

Figure~\ref{fig:efficiency} shows 0 order transmission efficiency of
a VPHG for various incidence angles at ambient temperature and
cooled down to $-180\degree$C. At both temperatures, the blaze is
still at 550~nm and the observed fluctuations are less than 5\%.
This value is of the order of the shift due to grating-stand thermal
contraction and errors due to the goniometric-arms repositioning. No
significant decrease of the efficiency has been seen.

\begin{figure}
  \subfigure[Wavefront at ambient temperature (265~K).]{
    \includegraphics[scale=0.75]{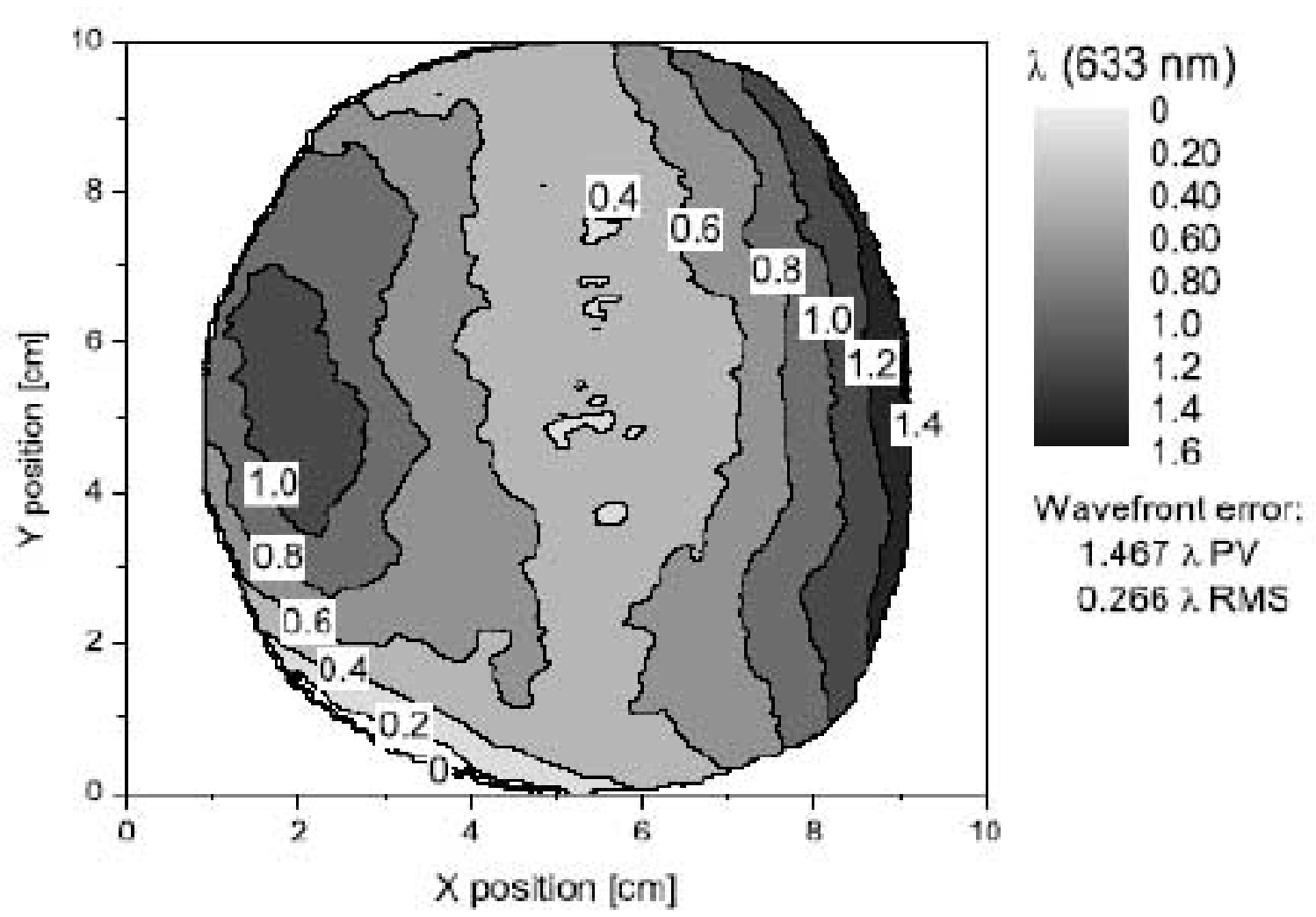}\label{fig:AmbiantWF}}\\
  \subfigure[Wavefront at 150~K.]{
    \includegraphics[scale=0.75]{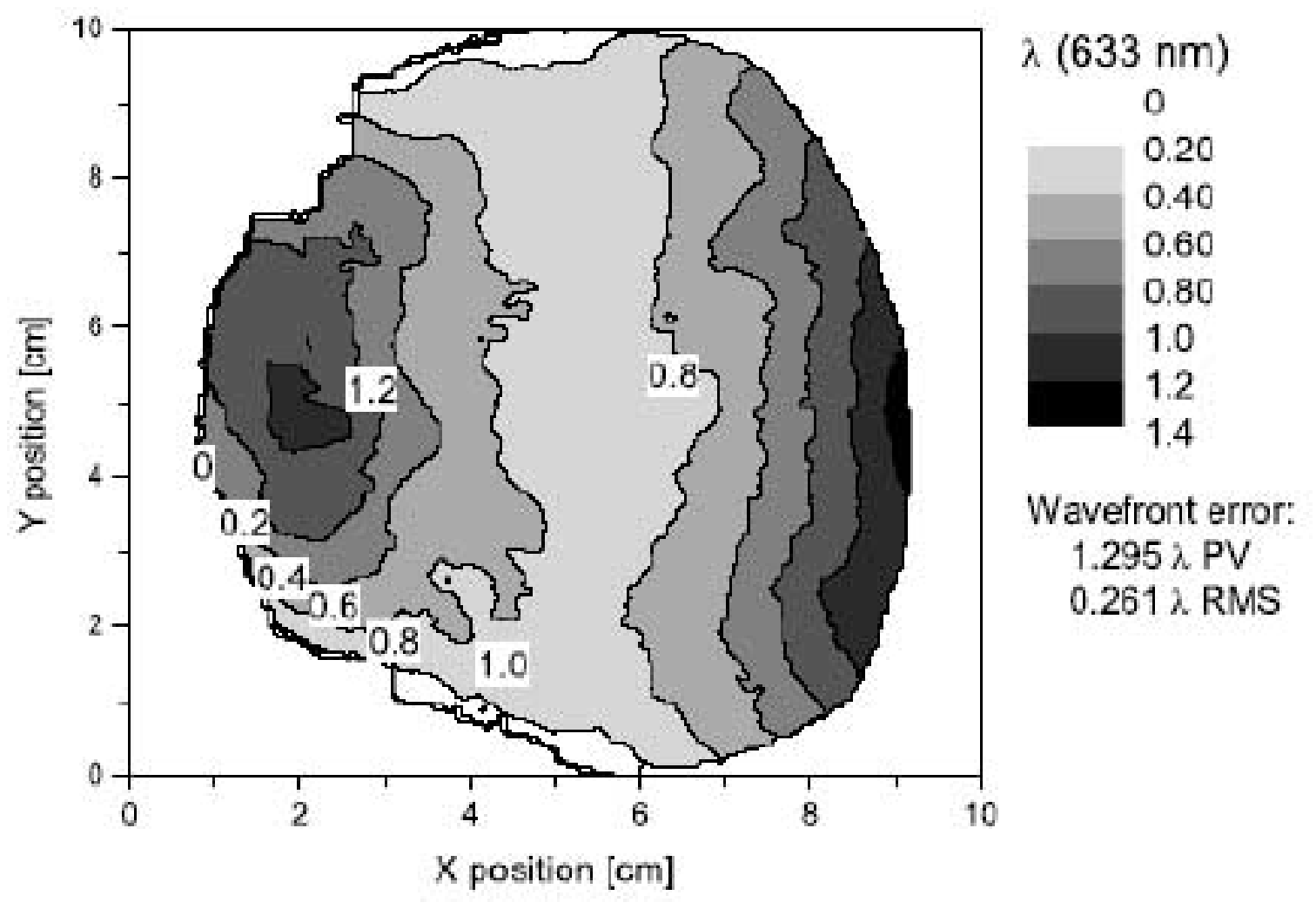}\label{fig:CryoWF}}
  \caption{Dif\-frac\-ted wavefront error recorded when the VPHG is at
  ambient (a)
           or cryogenic temperature (b).}
  \label{fig:cryowavefronts}
\end{figure}

\emph{Wavefront:} We measured the wavefront diffracted by one of our
VPHG at ambient temperature, during cool-down and at stable
cryogenic temperature. We notice that the higher the thermal
gradient on the grating, the larger is the wavefront error. But,
when thermal stability is ensured by few degrees, the wavefront
relaxes near to its original shape. This is what is depicted in
Fig.~\ref{fig:cryowavefronts}, where the diffracted wavefront errors
measured by a Zygo interferometer at ambient temperature and at
150~K are plotted. In both graphs, amplitude and shape of the
wavefront is similar.

It must be noted that all tested grating survived the thermal cycles
imposed on them. There were no degradation of the gelatine clarity
nor cosmetic damages.

\subsubsection{Post-polishing}

Depending on the blank thickness and characteristics, gelatine
processing and cement shrinkage can induce some stress. This stress
can lead to wavefront deformation. A solution is to correct the
wavefront by post-polishing after the grating has been encapsulated.
Because VPHGs are so sturdy this is feasible. However, the
diffracted wavefront can be a complex function without any symmetry.
In this case, classic polishing is rather tricky. ATHOL used the ion
beam figuring method (see Sect.~\ref{S_IBF}) where an ion gun throws
particles to the grating substrate and removes material where
needed.

\begin{figure}
      \subfigure[Before polishing.]{
      \includegraphics[scale=0.75]{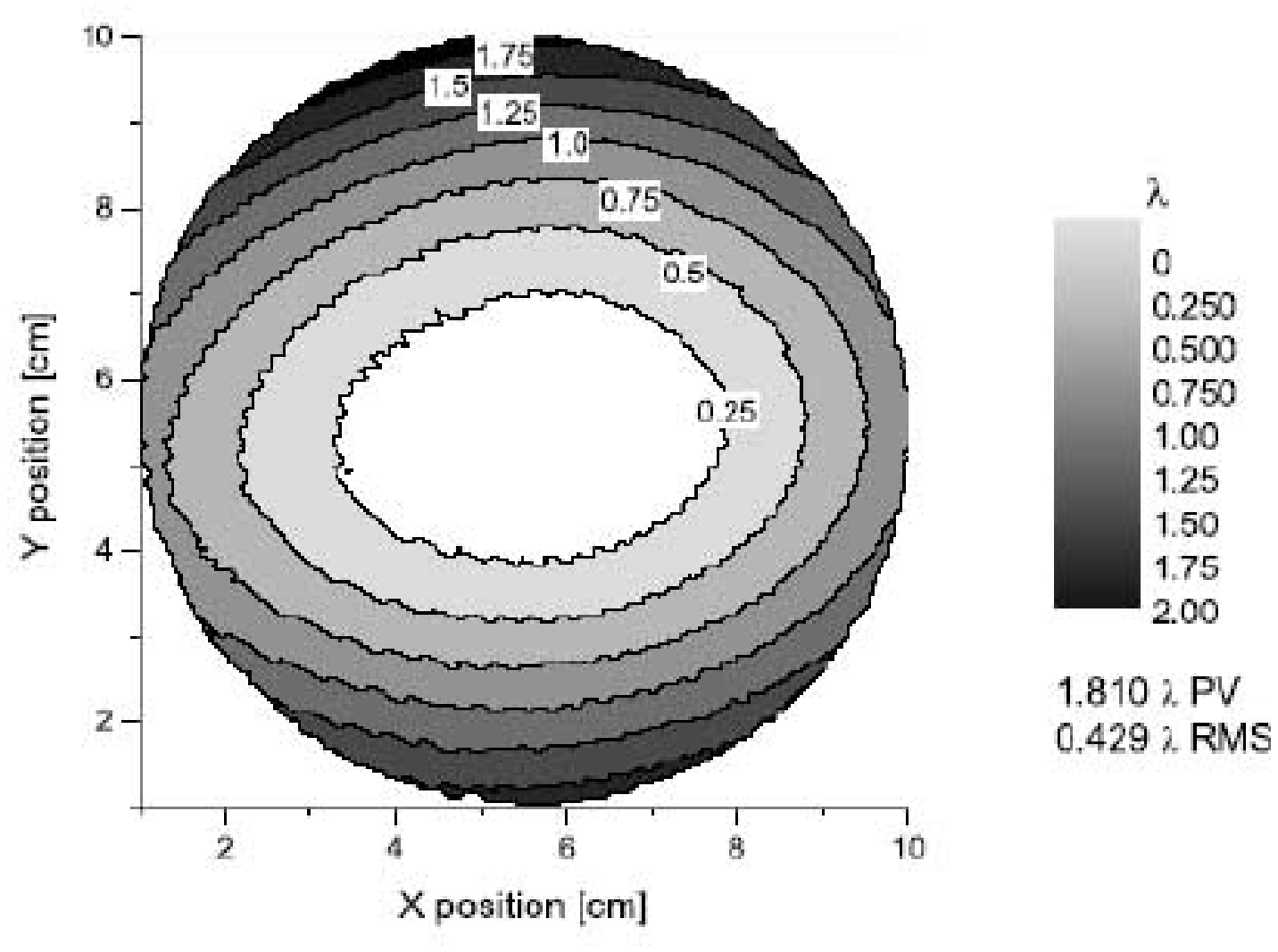}\label{fig:BeforeWF}}\\
      \subfigure[After ion beam figuring.]{
      \includegraphics[scale=0.75]{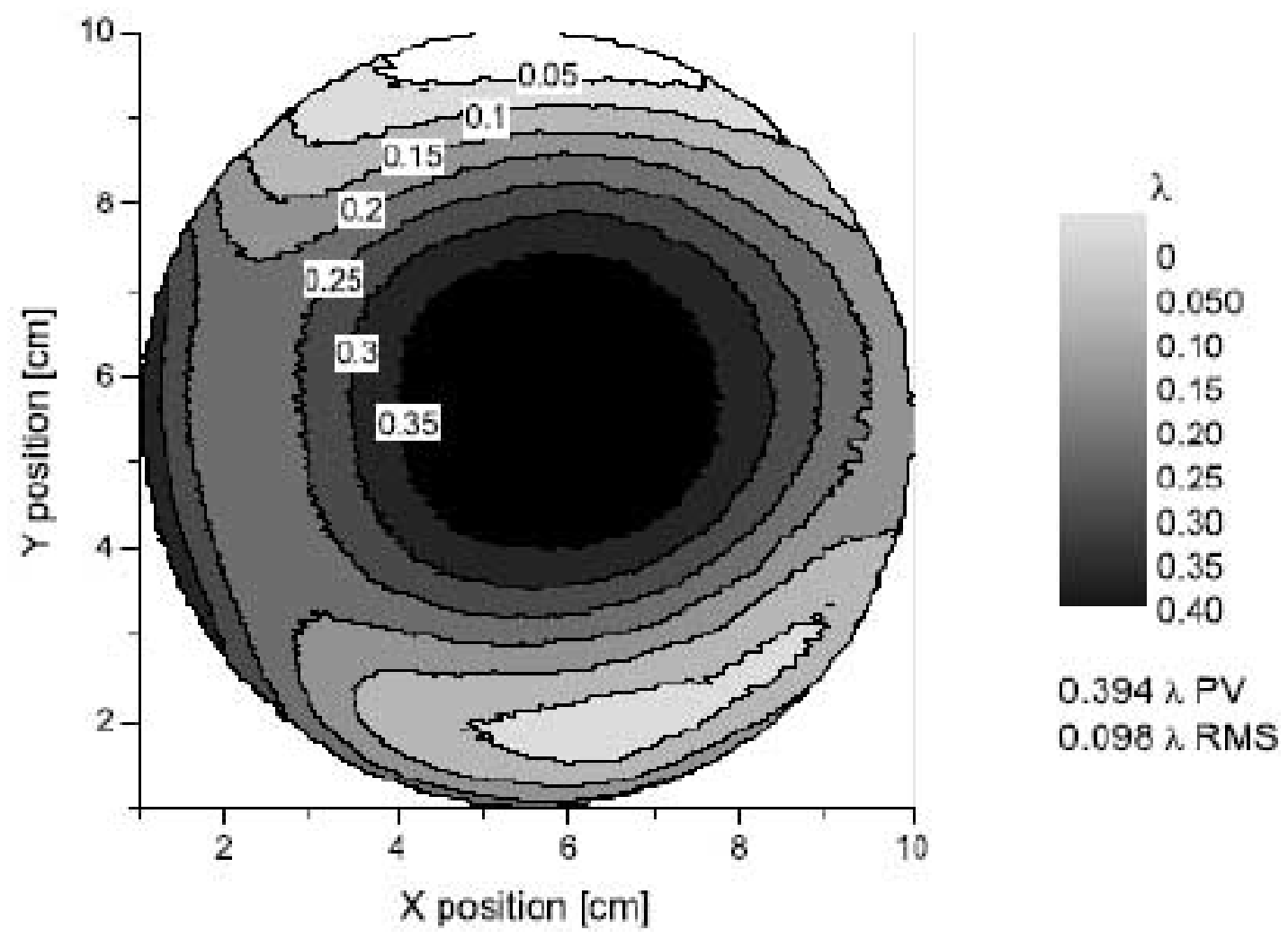}\label{fig:PolishWF}}
      \caption{Dif\-frac\-ted wavefront error from the VPHG before (a) and
      after (b) post-polishing.}
      \label{fig:post-polishing}
\end{figure}

Figure~\ref{fig:post-polishing} presents the diffracted wavefront of
a VPHG before and after post-polishing. The RMS wavefront error has
been reduced from $\lambda/2$ to $\lambda/10$ by ion beam figuring
over a diameter of 100~mm.

By using ion beam figuring it is also possible to induce any kind of
wavefront; converging, diverging, asymmetric, and so on. Of course,
optical functions can also be directly implemented into the hologram
during the recording.  Then, one can truly speak of a ``holographic
optical element''.

\subsection{Organic photochromic materials}

Photochromic materials change their absorption spectrum, and consequently their
color, by means of an optical stimulus. This change is reversible by using
light of different wavelength or it is simply induced by the temperature. This
kind of materials finds applications in many technological fields, for example
as sun filters, optical switches, optical memories (Crano \& Guglielmetti
1999a, 1999b). The two stable forms of photochromic materials differ not only
for color but also for refractive index, IR spectrum, redox potential and other
physical-chemical properties. Among the classes of photochromic materials, we
focused our attention and efforts to the class of diarylethenes, since they
have good properties in terms of high fatigue resistance, thermal stability,
and photochromic efficiency, which are important for practical application
(Irie 2000).

Usually the materials are used in solid state and in particular as films of
different thickness; therefore, it becomes important to have materials with
good filmable properties. By using low molecular weight molecules, a polymer
matrix (such as polymethylmetacrilate) is usually needed and this limits the
content of active molecules to few percent ($<$10\%) in order to keep the film
homogeneity. Another approach is the synthesis of backbone photochromic
polymers, which combine the processability of the polymer materials and the
optical properties of photochromic materials (Stellacci et al. 1999, Kim et al.
2002, Bertarelli et al. 2004, Wigglesworth et al. 2004). We used the
photochromic polymers to make rewritable elements for astronomical
instrumentation.

\subsubsection{Focal plane masks}

\begin{figure}
\resizebox{\hsize}{!} {\includegraphics[]{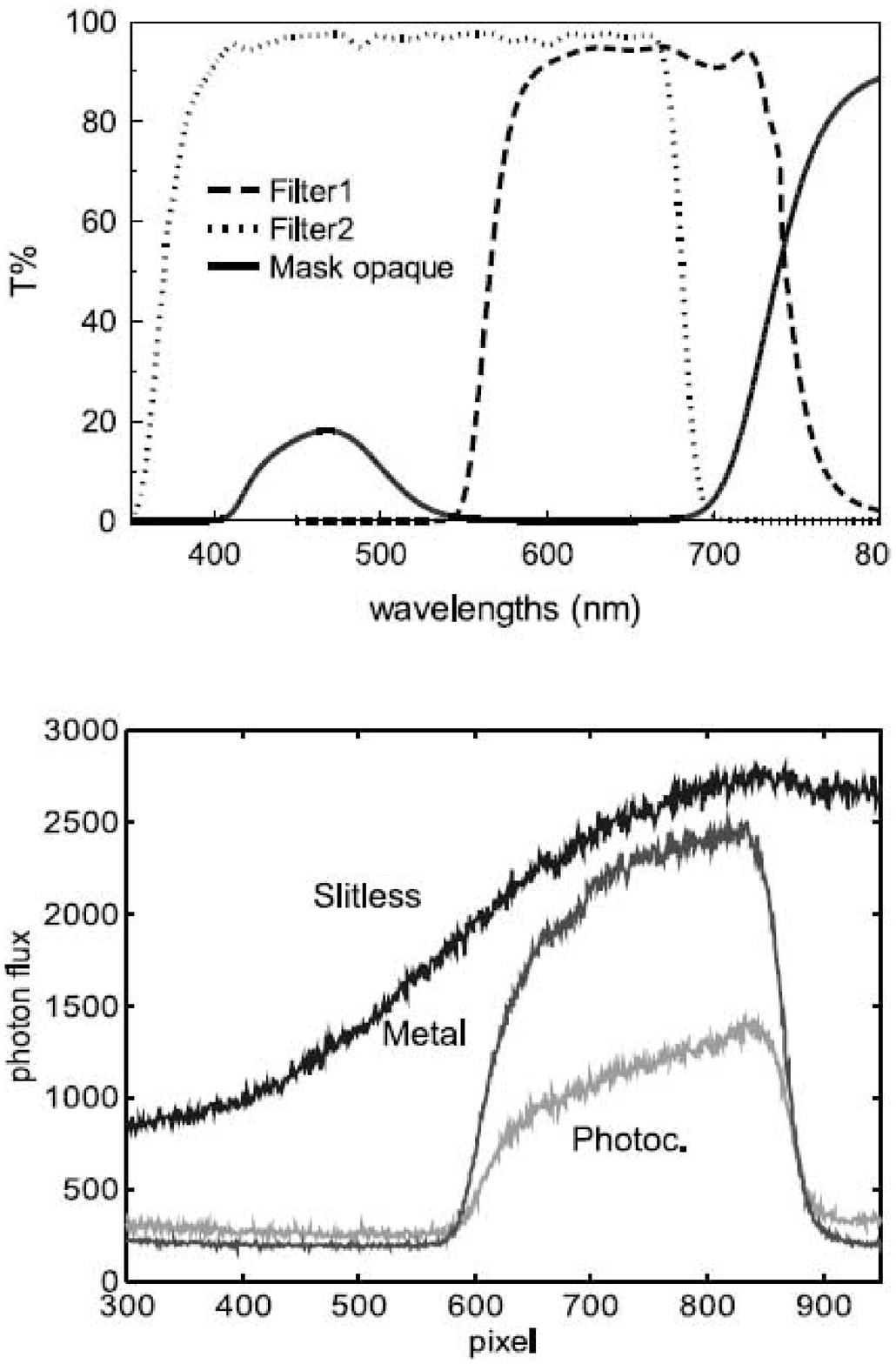}}
\caption{(Top:) Transmission curves of an opaque photochromic FPM
and of the filters. (Bottom:) CCD trace of a metallic mask, a
photochromic mask and the slitless case. } \label{fig:bianco1}
\end{figure}

We exploited the change in transparency of a photochromic film in the visible
to make rewritable focal plane MOS masks (Multi Object Slits; Molinari et al.
2002, Bianco et al. 2005): the film is turned into the opaque form, then a red
laser writes the slits turning the  photochromic film into the transparent
form. OAB built up a complete set-up to read the pre-imaging and, after
defining the slits position, to write automatically the slit pattern. The
spectral range is limited (500--700 nm) by the photochromic material and the
filters used. The performances can be improved by choosing proper filters that
match exactly the passband of the photochromic film.
Figure~\ref{fig:bianco1}\emph{(top)} shows a plot of the transmission curves of
a photochromic mask (70$\mu$m thick) in the opaque form and the curves of two
filters designed to match the film passband. In this configuration, the whole
system is opaque over the entire range of sensitivity of the Si CCD.

The most important parameter related to the transmission curves of the mask is
the contrast between the slits and the mask. Since the photochromic  film is
not completely opaque, there is a background noise that is not present in the
metallic mask, especially in the spectroscopic mode. The results are shown in
Fig.~\ref{fig:bianco1}\emph{(bottom)}, obtained with the AFOSC camera (Asiago
telescope, Italy) by  using a flat field as light source and a single slit. The
metallic mask clearly shows an higher signal through the slit compared to the
photochromic mask. The ratio between the signals is about two and consequently
the noise of the metallic mask is smaller.

\subsubsection{Volume phase holographic gratings}

\begin{figure}
\resizebox{\hsize}{!}{\includegraphics[]{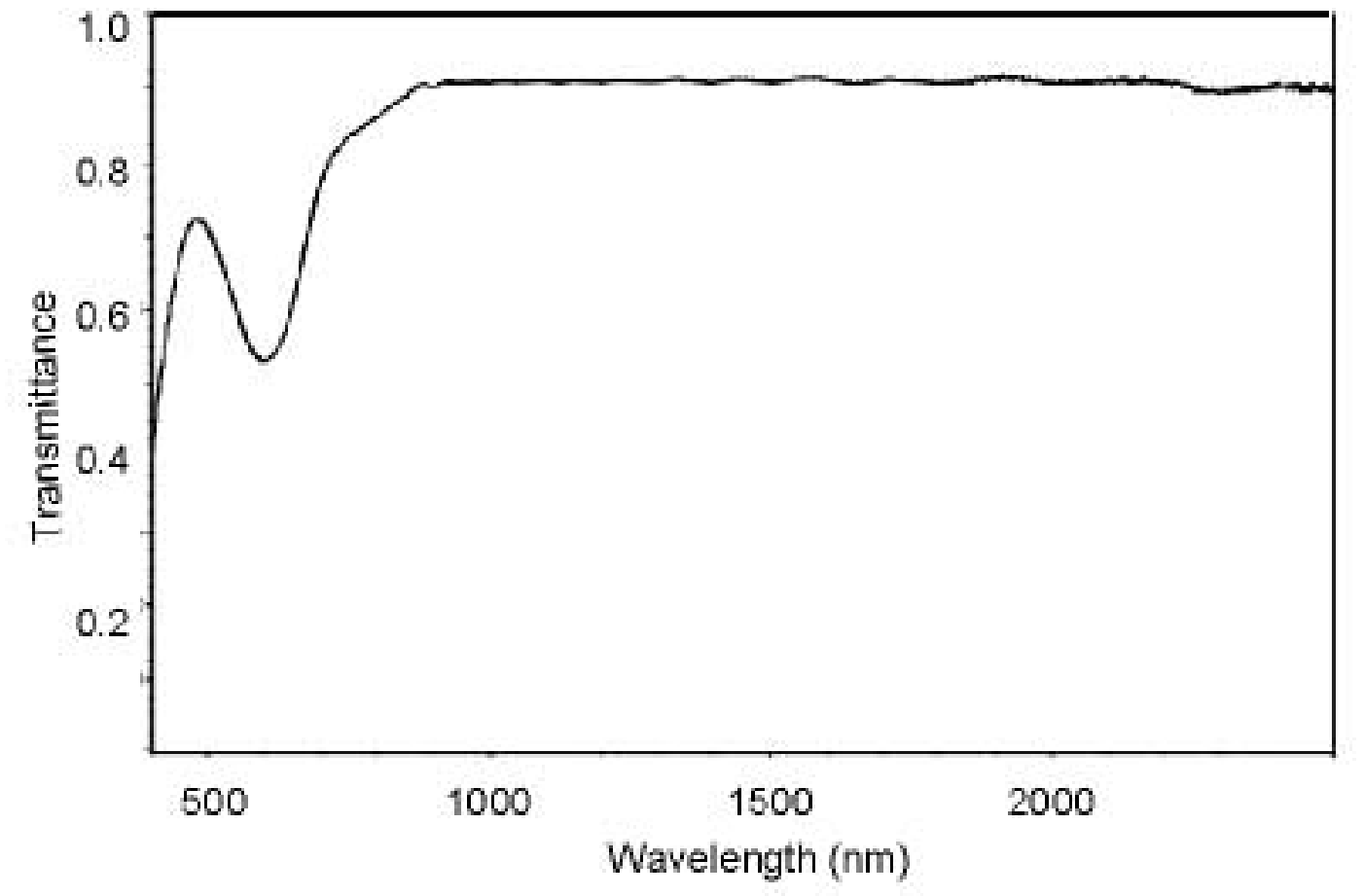}}
\caption{Transmission curve of a film of a photochromic polyester (thickness 5
microns).} \label{fig:bianco2}
\end{figure}

Efficient dispersing elements are of great interest in astronomical
instrumentation; moreover, the possibility of making large size gratings fits
the needs of new instrumentations for current large telescopes, and the next
generation ELTs. Volume phase holographic gratings (VPHGs) are good candidate
for these aims (Barden et al. 1998, Barden et al. 2000, Blanche et al. 2004).
This kind of dispersing element exploits the periodical difference in
refractive index that takes place in particular materials, therefore the
diffraction happens in the volume of the material. Usually, VPHGs are made of a
thin layer of dichromated gelatine (DCG), which is cast on a glass substrate,
exposed to laser light pattern (488 nm) and then developed by a chemical
treatment (Curran \& Shankoff 1970; see also Sect.~\ref{VPH_blanche}). The
efficiency curve of these gratings depends on the thickness of the layer and on
the modulation of the refractive index ($\Delta n$). Because of the developing
process, the thickness of DCG layer is limited to 20–30 microns. By using a
material that changes the refractive index without any chemical process remove
this limit and it makes easier the VPHG manufacturing. Photochromic materials
show a $\Delta n$ between the two forms and this difference is easily achieved
by using light of suitable wavelength (Bertarelli et al. 2004), since the
change in refractive index is due to a different electronic structure of the
molecules instead of a different density as in the case of DCG. The useful
spectral range for VPHGs made of photochromic materials (Molinari et al. 2002)
is the near infrared where the material is completely transparent (see
Fig.~\ref{fig:bianco2}); the transmittance is close to 90\% between 800 nm and
2500 nm, which means, taking into account the reflection losses (more than
8\%), a negligible absorption or scattering. These VPHGs will be useful for
astronomical instrumentations working in the J, K, L bands.

\begin{figure}
\resizebox{\hsize}{!}{\includegraphics[]{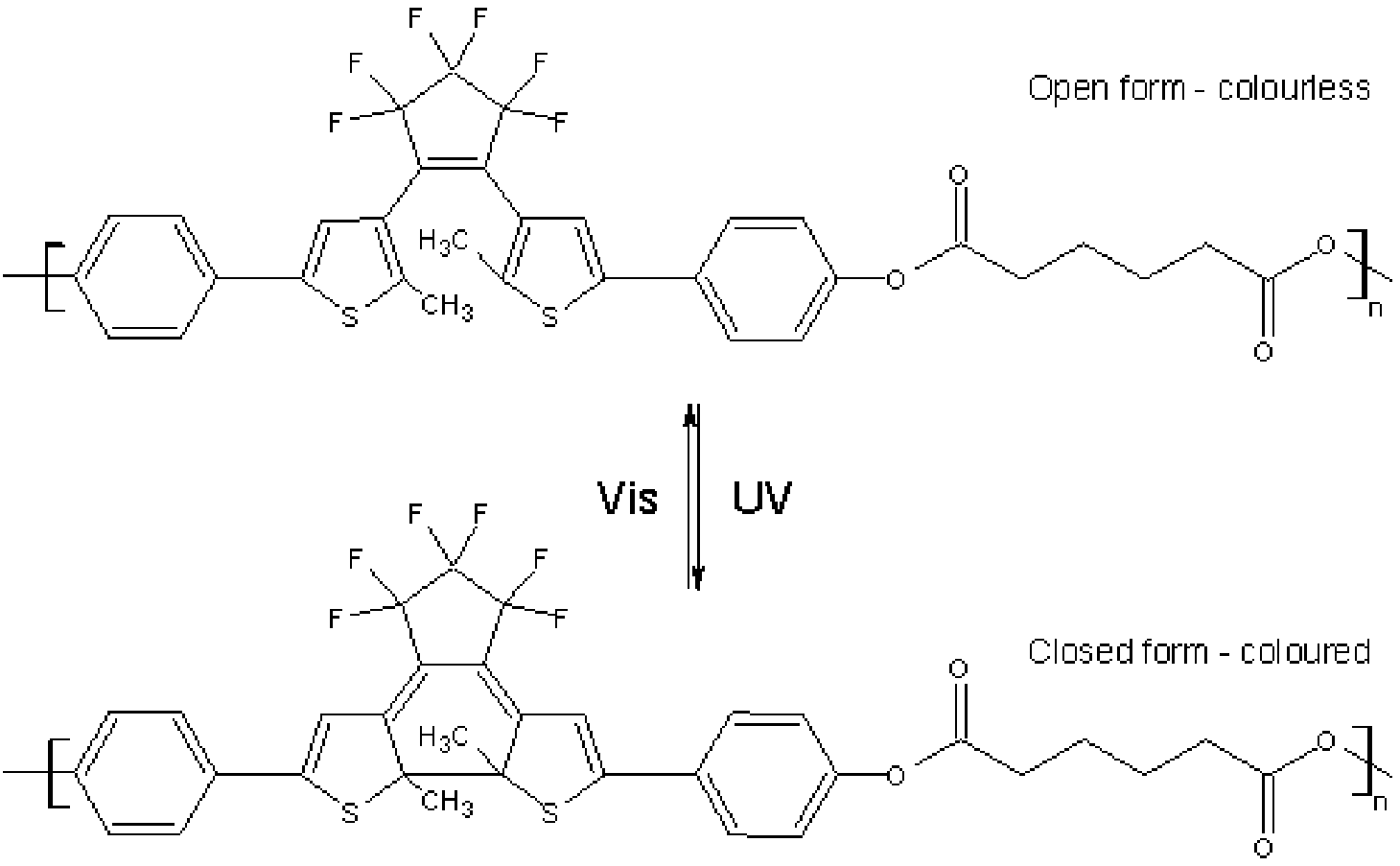}}
\resizebox{\hsize}{!}{\includegraphics[]{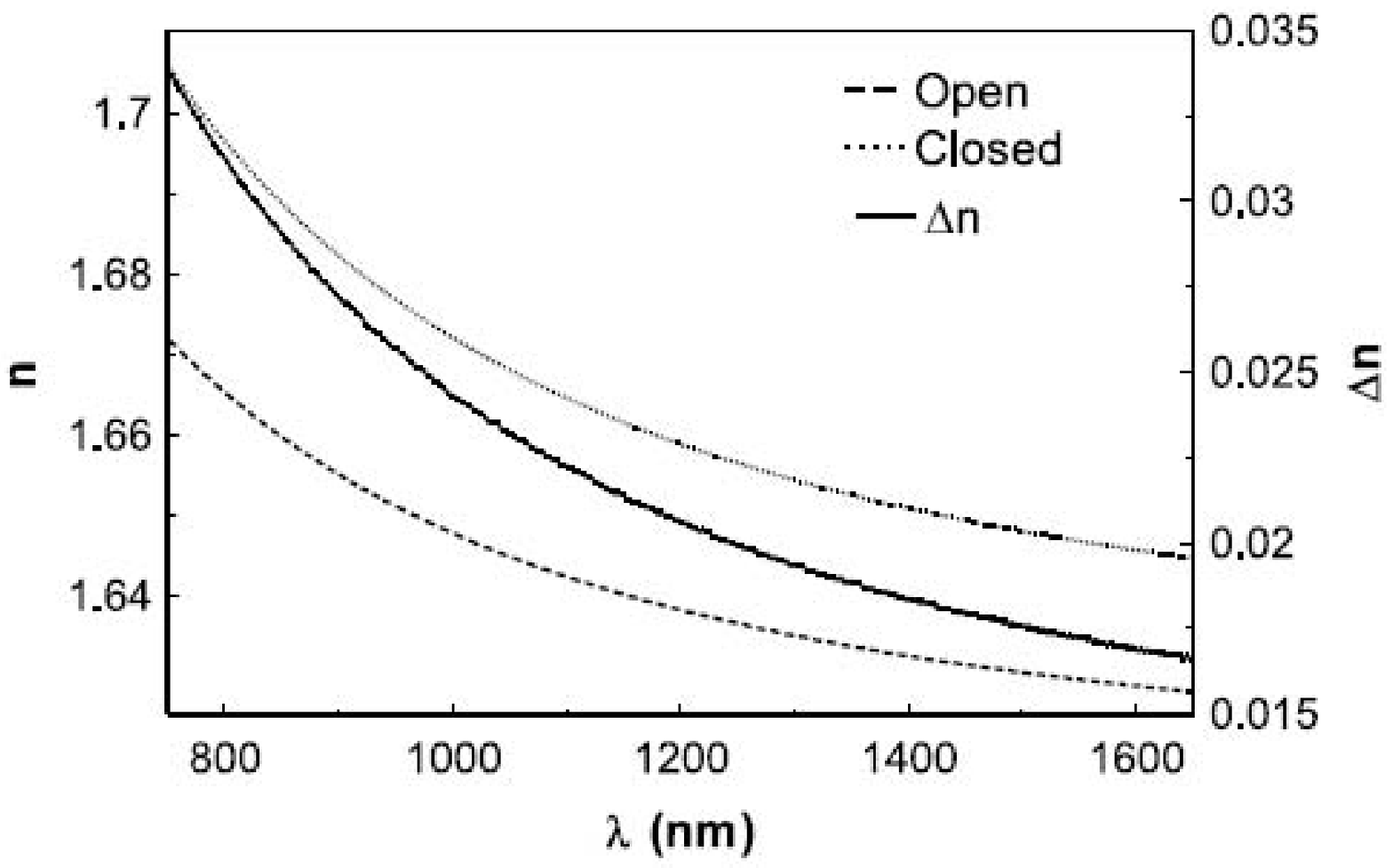}} \caption{(Top:) Chemical
structure of polyester P1 in the open and closed forms. (Bottom:) The
refractive indices $n$ and $\Delta n$ curves  of the two forms of P1.}
\label{fig:bianco3}
\end{figure}

Reaching large $\Delta$n for films of photochromic materials ($\sim$0.01--0.08)
is the fundamental requirement in order to make VPHGs with large peak
efficiency and without a shrinkage of the blaze curve (increasing the
thickness). Table~\ref{tab:bianco1} clearly shows that the $\Delta$n obtained
with low molecular weight photochromic diarylethenes is small, since they need
a polymer matrix, whereas the diarylethenes polymers have a large modulation.
Among the different classes of photochromic polymers, polyester P1 (chemical
structure in Fig.~\ref{fig:bianco3}) was synthesized since they have good
efficiency in both the photochromic reactions; moreover, the molecular weight
is large enough to make make good optical films by casting.

\begin{table}[h]
\caption{Refractive index modulation measured for different photochromic
materials. } \label{tab:bianco1}
\begin{tabular}{ll}\hline
Photochromic materials  &  $\Delta n$ (1.5 $\mu$m) \\ \hline
Low molecular weight molecules (a)  &  0.001--0.005 \\
Backbone photochromic polymers (b)  &  0.003--0.032 \\
\\\hline
\end{tabular}
\end{table}

The refractive index of the two forms of the photochromic polyesters was
measured by using spectral reflectance between 800 and 1600 nm. A thin film
(300 nm) was spin coated on a glass substrate and the reflectance is measured.
A fit based on a Cauchy model ($n = A+B/\lambda ^2 + C/\lambda ^4$) was applied
and both the thickness and the refractive index were determined. The maerial in
its coloured form has a refractive index larger  than the colourless form. The
$\Delta$n increases when the wavelength decreases. This is a difference from
the DCG, which shows a constant modulation, and it affects the efficiency curve
of the VPHG. At 1500 nm the $\Delta n$ of P1 is 0.0165, large enough for an
efficient VPHG (see Fig.~\ref{fig:bianco3}).

The sensitivity of P1 was also measured by using green laser (514 nm) starting
from a film in the coloured form. The laser power and the area of the spot were
measured, then the polyester film was lightened with the laser and the laser
power was monitored until a plateau was reached as function of time. At the
wavelength used the sensitivity is 120 J/mm$^3$. This parameter is important to
determine the exposure time of the film to the light pattern. Films of pure P1
for VPHGs were obtained by using a control coater with a spreading blade, that
casts the polymer solution at constant thickness. The target thickness was
20$\mu$m, but homogeneous films of only 5 microns were cast successfully. Low
viscosity of the solution was most probably the cause of the too thin films.
Some attempts to increase viscosity were made by adding a small amount of PMMA
with an high molecular weight (120,000 or 1,000,000 g/mol), but in this case,
the films turned out to be opaque. Consequently the efficiency of a VPHG could
be very low. Starting from a film 4.5$\mu$m thick, a VPHG was written by
transferring the pattern of a Ronchi ruling glass slide (600 l/mm, OD=3) with a
green laser (532nm, 30mW).

\begin{figure*}
\resizebox{\hsize}{!} {\includegraphics[]{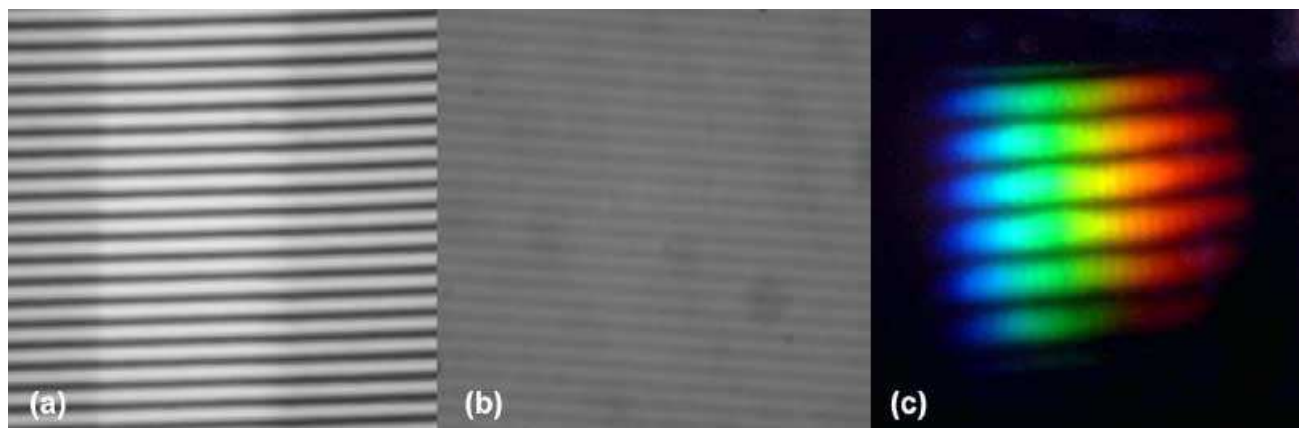}} \caption{Optical
microscope images of the 600 l/mm pattern: (a) the Ronchi glass slide; (b) the
photochromic film; (c) the diffraction pattern of a white light lamp.}
\label{fig:bianco4}
\end{figure*}

In order to check the result of the writing procedure, images of the recorded film were
obtained by optical microscopy. Figure~\ref{fig:bianco4}(a,b) shows the
similarity of the pattern of the Ronchi glass slide and the photochromic film.

At this level it was not possible to quantify the contrast profile of the
pattern, which could give the conversion grade of the photochromic layer, but
we were able to obtain a diffraction pattern from a white lamp and our
photochromic grating (see Fig.~\ref{fig:bianco4}c). At present, the main issue
remains the possibility to make thick films based on photochromic polymers.

\begin{figure}
\resizebox{\hsize}{!} {\includegraphics[]{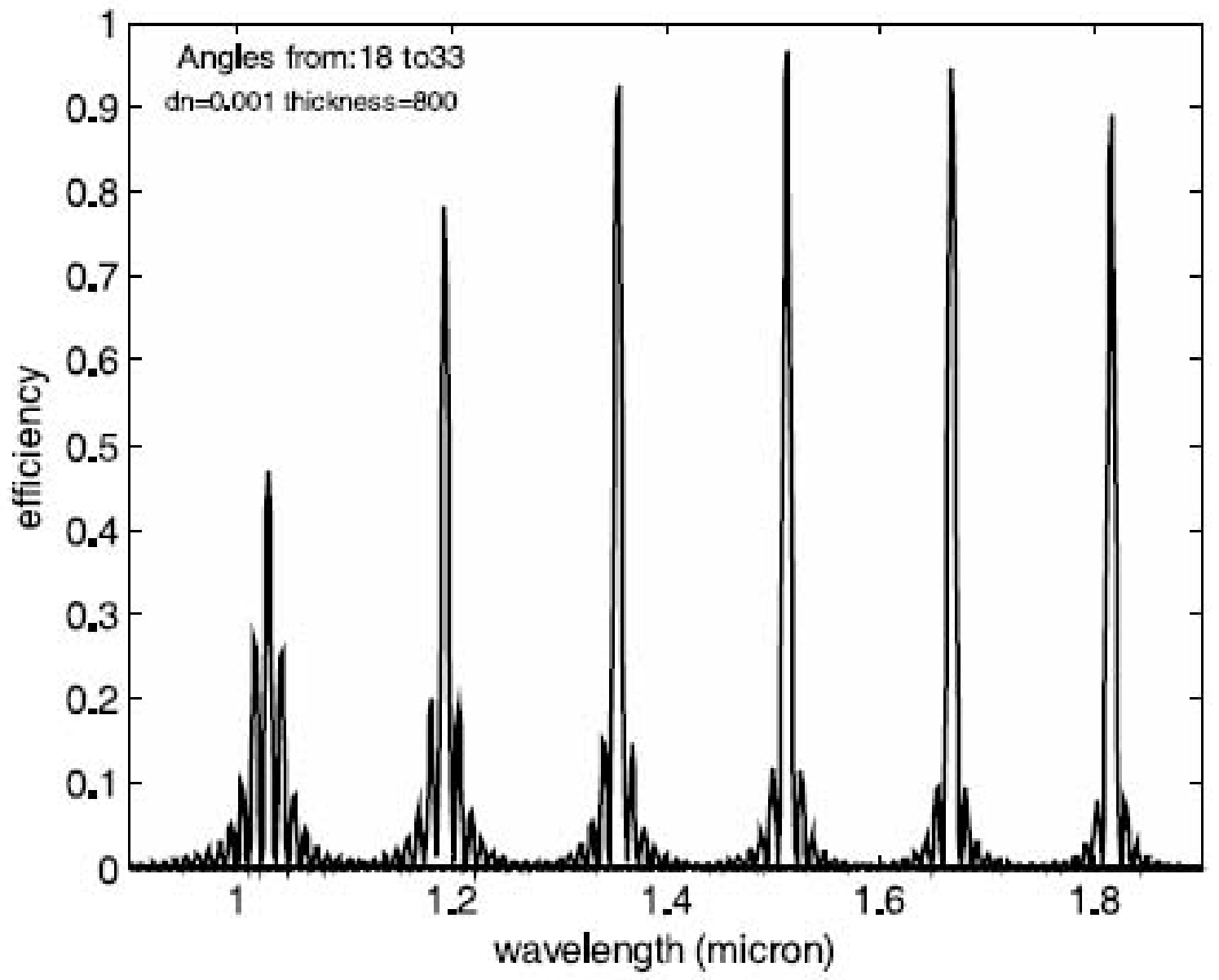}}
\caption{Efficiency curves of a VPHG with 600 l/mm for different
incident angles, based on a Rigorous Coupled Wave Analysis (RCWA)
simulation.} \label{fig:bianco5}
\end{figure}

Usually the dispersing elements must show high efficiency over a wide spectral
range; for VPHGs it means a large refractive index modulation and thin films.
Increasing the thickness and lowering the $\Delta$n, the passband of the VPHG
shrinks and the peak efficiency remains close to 1. In this configuration the
grating behaves as a filter with a narrow passband that can be tuned in
wavelength by changing the incidence angle (Molinari et al. 2004, Havermeyer et
al. 2004). This kind of tunable filters could be robust and cheap. Simulations
based on Rigorous Coupled Wave Analysis (RCWA, Gaylord \& Moharam 1982) were
carried out on a thick photochromic film (800 microns,$\Delta$n = 0.001) to
verify this possibility and the results are summarized in
Fig.~\ref{fig:bianco5}.

\begin{figure}
\resizebox{\hsize}{!} {\includegraphics[]{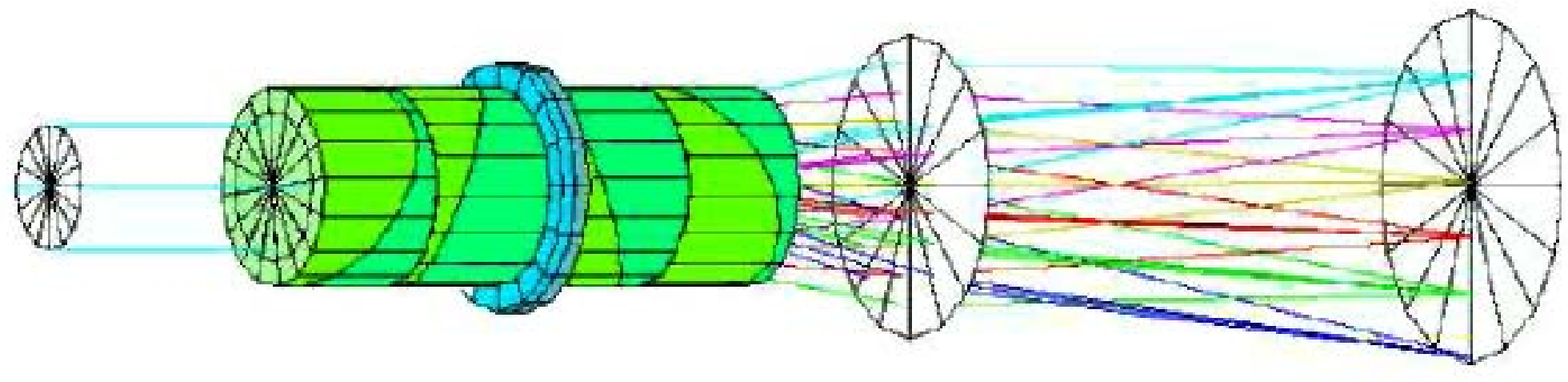}}
\caption{Optical configuration for a tunable filter based on a VPHG
with narrow passband. The two counter-rotating Rayleigh prisms that
modulate the incidence angle on the grating are shown in the
middle.} \label{fig:bianco6}
\end{figure}

A thick film of optical polymers such as polymethylmethacrylate (PMMA) doped
with photochromic molecules can be a good substrate, since, as mentioned
previously, the photochromic materials do not need a chemical process. A simple
device which exploits the features of such VPHGs is composed of two
counter-rotating Rayleigh prisms which modulate the incidence angle on the
grating leaving the optical axis of the system unperturbed (see
Fig.~\ref{fig:bianco6}).

\subsection{Sintered SiC mirrors for ELTs}

The recurring concept for ELT's outstandingly large mirrors (primary
and also secondary) is using a hexagonal segmentation, with a flat
to flat size between 1m and 3m. The main drivers for the choice of
the mirror segment material are:
\begin{itemize}
  \item Specific stiffness and thermal stability and
  \item manufacturing capability at cost effective condition and with
  reasonable time span.
\end{itemize}

\subsubsection{SiC material characteristics}

The {\sl Boostec} material is a sintered silicon carbide (SSiC)
which has been fully characterized, even down to cryogenic
temperatures. In comparison with the glass-ceramics and the other
types of silicon carbide materials, it offers a very interesting
package of properties. These SiC materials are:
\begin{itemize}
      \item SiSiC, reaction bonded or Si infiltrated,
      \item C/SiC or CeSiC, including short carbon fibres, also Si
            infiltrated, and
      \item CVD (Chemical Vapour Deposited) SiC.
\end{itemize}

\noindent The telescope materials are selected for
\begin{itemize}
      \item their high specific stiffness (Young's modulus / density) which
            enables manufacture of lightweight parts with high mechanical
            stability;
      \item their high thermal stability (thermal conductivity / coefficient
            of thermal expansion), which give a lack of sensitivity
            to temperature changes.
\end{itemize}

\begin{figure}
\resizebox{\hsize}{!} {\includegraphics[]{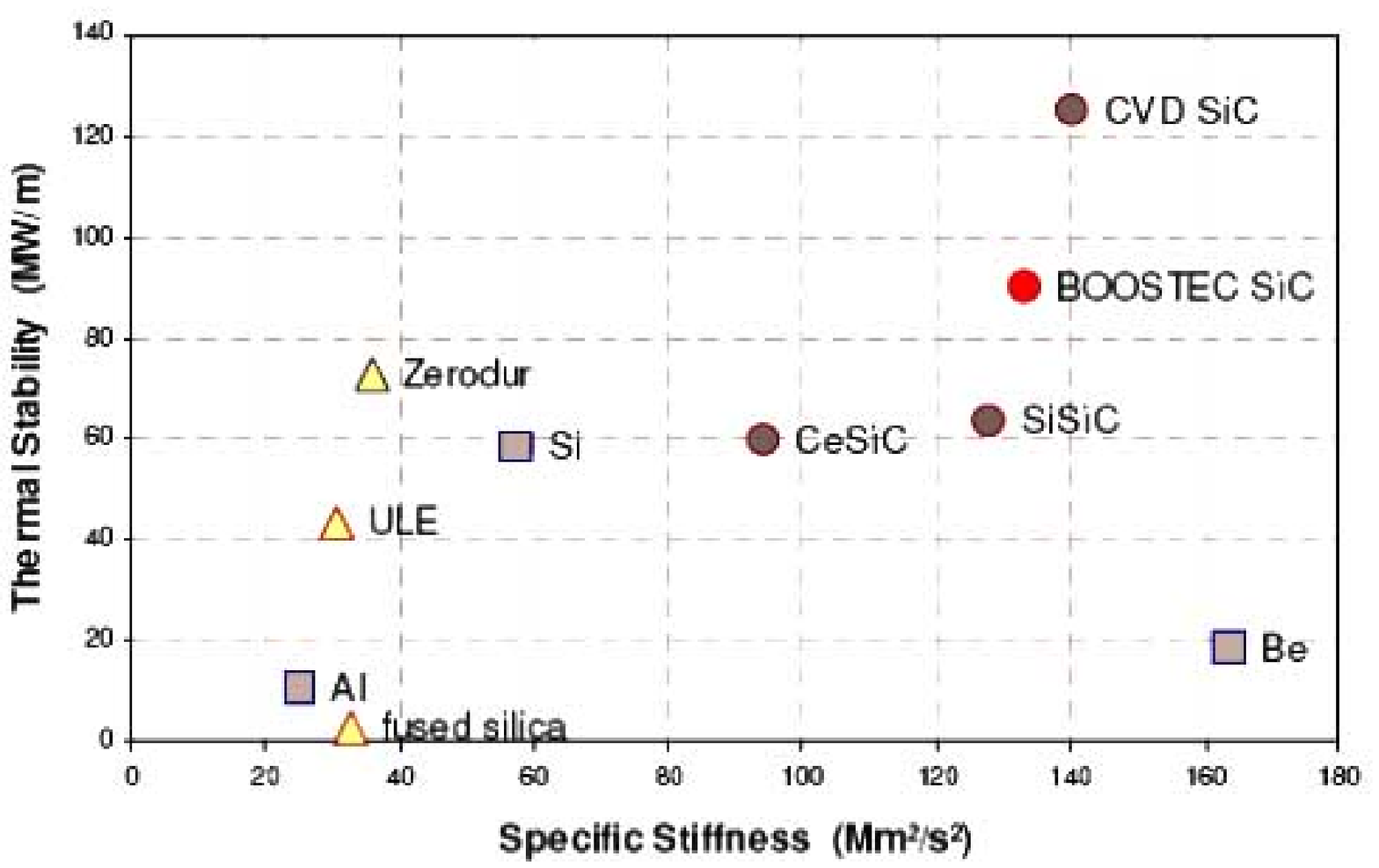}}
\caption{Figures of merit for telescope material choice.}
\label{fig:bougoin1}
\end{figure}

Thanks to its high stiffness (420 GPa) and its high thermal
conductivity (180~W/mK at RT) the {\sl Boostec} SiC appears among
the best choices in Fig.~\ref{fig:bougoin1}. The only CVD SiC shows
better properties, thanks to both a very high purity and a total
lack of porosity. Unfortunately, all attempts for manufacturing
large monolithic mirror blanks with this CVD material failed. On the
other hand, its physical properties fit very well with the ones of
the sintered SiC; this is the reason why it is currently suggested
by {\sl Boostec} as a cladding on its material, when a small surface
porosity cannot be tolerated. The Figures of Merit of {\sl Boostec}
SiC are enhanced when the temperature is decreased from room
temperature to 100~K.

\begin{figure}
\resizebox{\hsize}{!} {\includegraphics[]{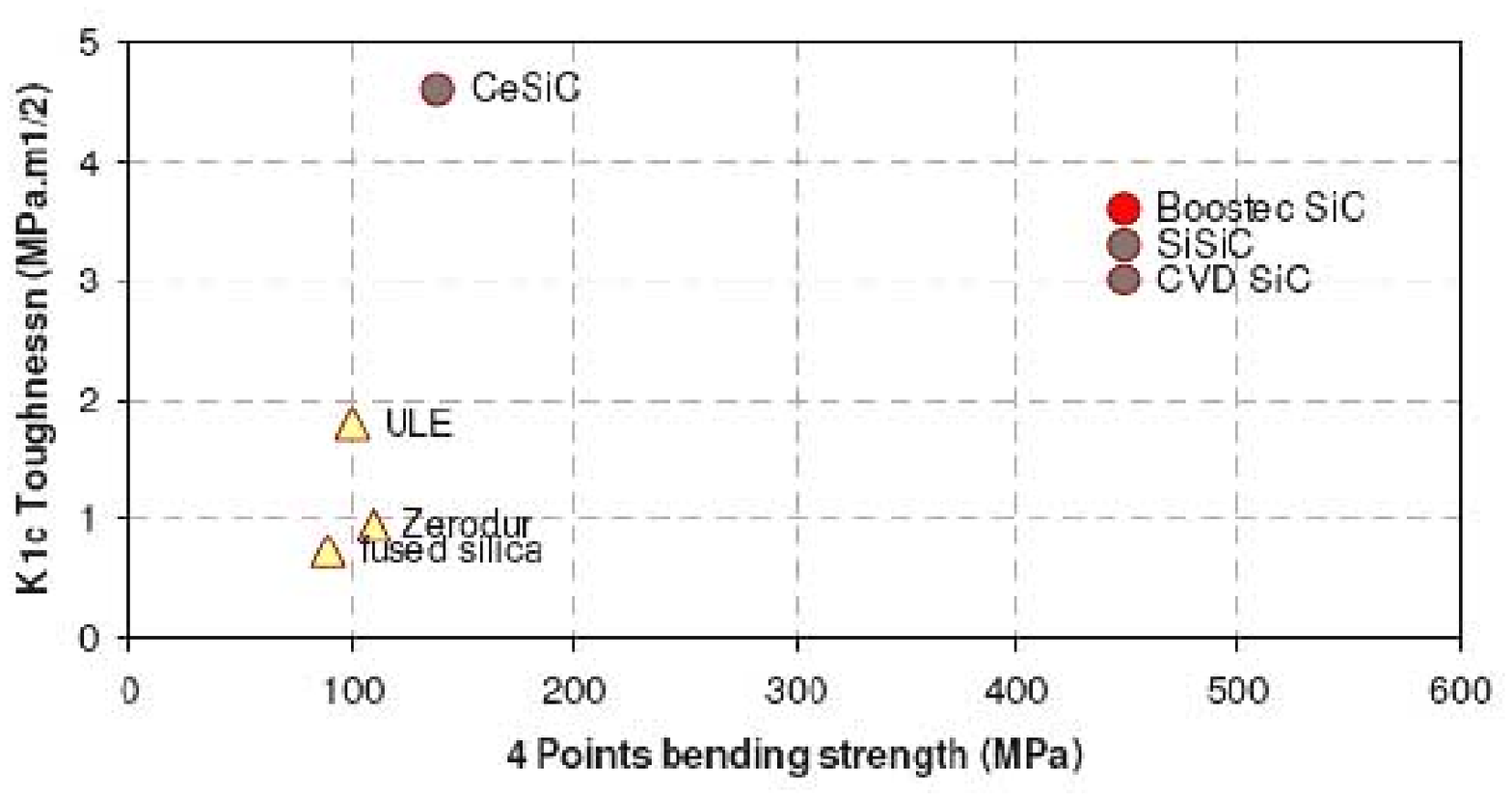}} \caption{Mechanical
strength and toughness.} \label{fig:bougoin2}
\end{figure}

All SiC materials exhibit significantly higher strength and
toughness than fused silica and also than glass-ceramics
(Fig.~\ref{fig:bougoin2}). An addition of short carbon fibres
slightly improves the K1c toughness which can be assumed as the
resistance to the crack propagation. But this is clearly obtained to
the detriment of the mechanical strength of these C/SiC materials.
{\sl Boostec}'s sintered SiC, as well as infiltrated (fine
microstructure only) and CVD ones show the best mechanical strength.
Both CVD and {\sl Boostec} SiC exhibit homogeneous and isotropic
microstructure. They have no secondary phase, no outgassing, no
sensitivity to moisture and an outstanding resistance to strong
acids or alkalis.

{\sl Boostec} SiC is furthermore insensitive to mechanical fatigue.
It shows around 1.5 vol.\% closed porosity, which is commonly masked
with a CVD SiC cladding when necessary. Even if they are harder than
the glass-ceramics, {\sl Boostec} SiC or its possible SiC CVD
cladding can be easily polished with standard mechanical diamond
polishing and Ion Beam Figuring. Their single phase, their fully
isotropic properties, their lack of plasticity and their high
stiffness are very helpful. They do not require any relaxation time
thanks to both their low mass and high thermal conductivity. All the
standard optical coatings of glass-ceramics can be also applied on
these SiC materials.

\subsubsection{SiC manufacturing capabilities and large size abilities}

\begin{figure}
\resizebox{\hsize}{!} {\includegraphics[]{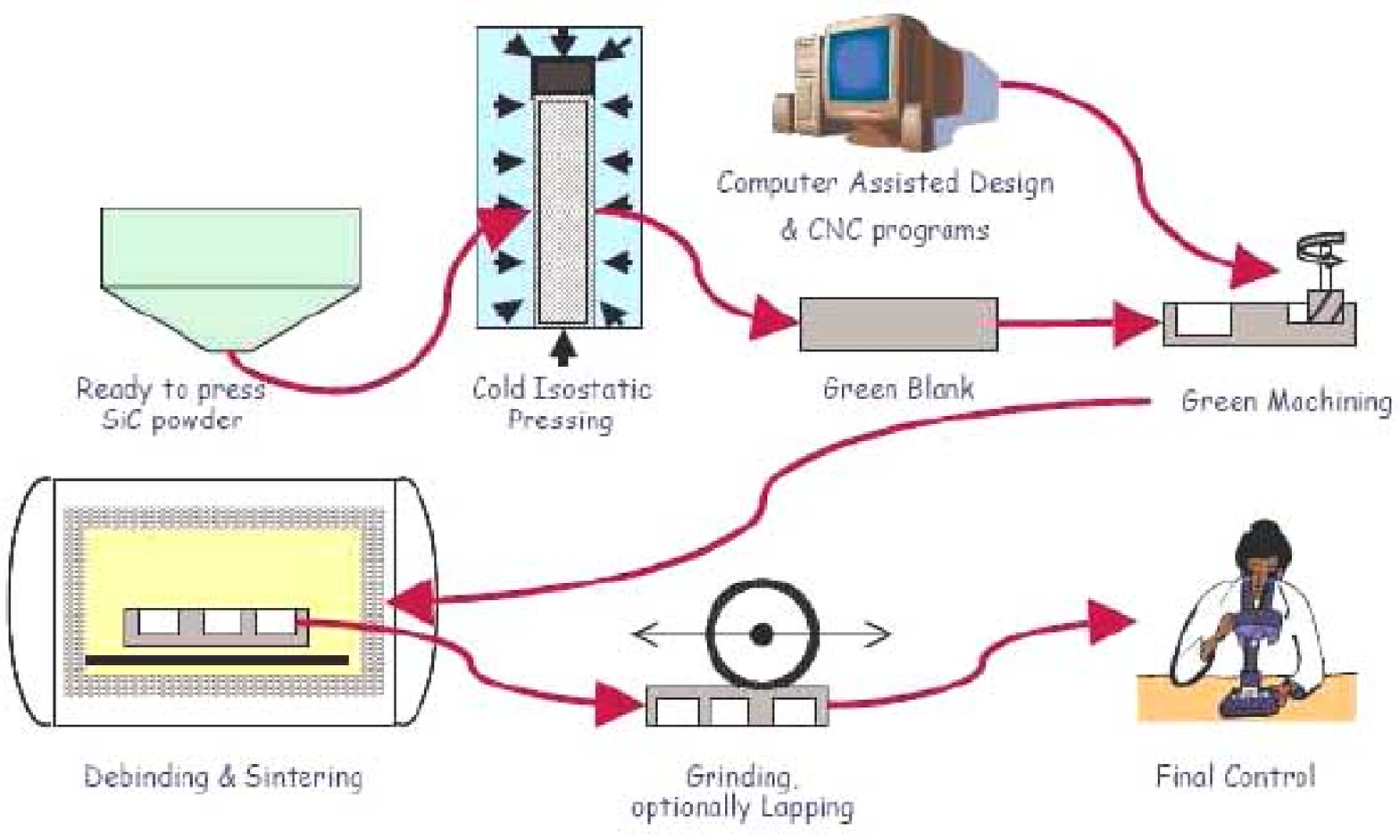}}
\caption{Manufacturing process of {\sl Boostec} SiC mirror blanks.}
\label{fig:bougoin3}
\end{figure}

\emph{Boostec SiC manufacturing process for monolithic parts:}
Monolithic SiC parts of up to 1.5x1.0m can be manufactured according
to a process which is summarized in Fig.~\ref{fig:bougoin3}. {\sl
Boostec} uses a near-net-shaping technique, i.e. that the shape of
the part is obtained mainly by machining the ``green blank'', before
sintering. Soft blanks machining makes {\sl Boostec} process very
productive and cost effective. All curved profiles, possibly
aspheric or off-axis, are obtained from CNC grinding machines, after
sintering.

\emph{Manufacturing very large SiC parts:} Sintered SiC parts can be
joined together thus obtaining parts larger than 1.5x1.0m. To this
extend, {\sl Boostec} has developed and qualified a non reactive
brazing process. A braze alloy is used since its CTE fits very well
with that of SiC. This process has been successfully used on several
space projects such as the 1.5-m diameter parabola of the {\sl
Aladin} telescope and the 3.5-m diameter {\sl Herschel} primary
mirror. No technological limitation has been revealed so that sizes
limits of both monolithic and brazed blanks could most likely be
increased.

\emph{Assembling SiC parts:} Two assembling techniques have been
developed, qualified and also successfully used in space telescopes:
(a) bolting SiC-SiC or SiC-metal with metallic bolts, (b) gluing
SiC-SiC or SiC-metal with epoxy material.

\subsubsection{Production experience}

\emph{Mass production of sintered SiC:} The world production of
synthetic SiC is 900,000 tons per year from which a very small part
is used as raw material for the production of around 450 tons of
sintered SiC. The ability to manufacture very cost effective
components has already been demonstrated for highly competitive
areas like automotive or armours. Before developing SiC optics, {\sl
Boostec} team has accumulated more than 10 years experience in mass
production of seal or bearing rings made of sintered SiC. The
telescope material is the same as the one manufactured for such
industrial applications.

\begin{figure}
\resizebox{\hsize}{!} {\includegraphics[]{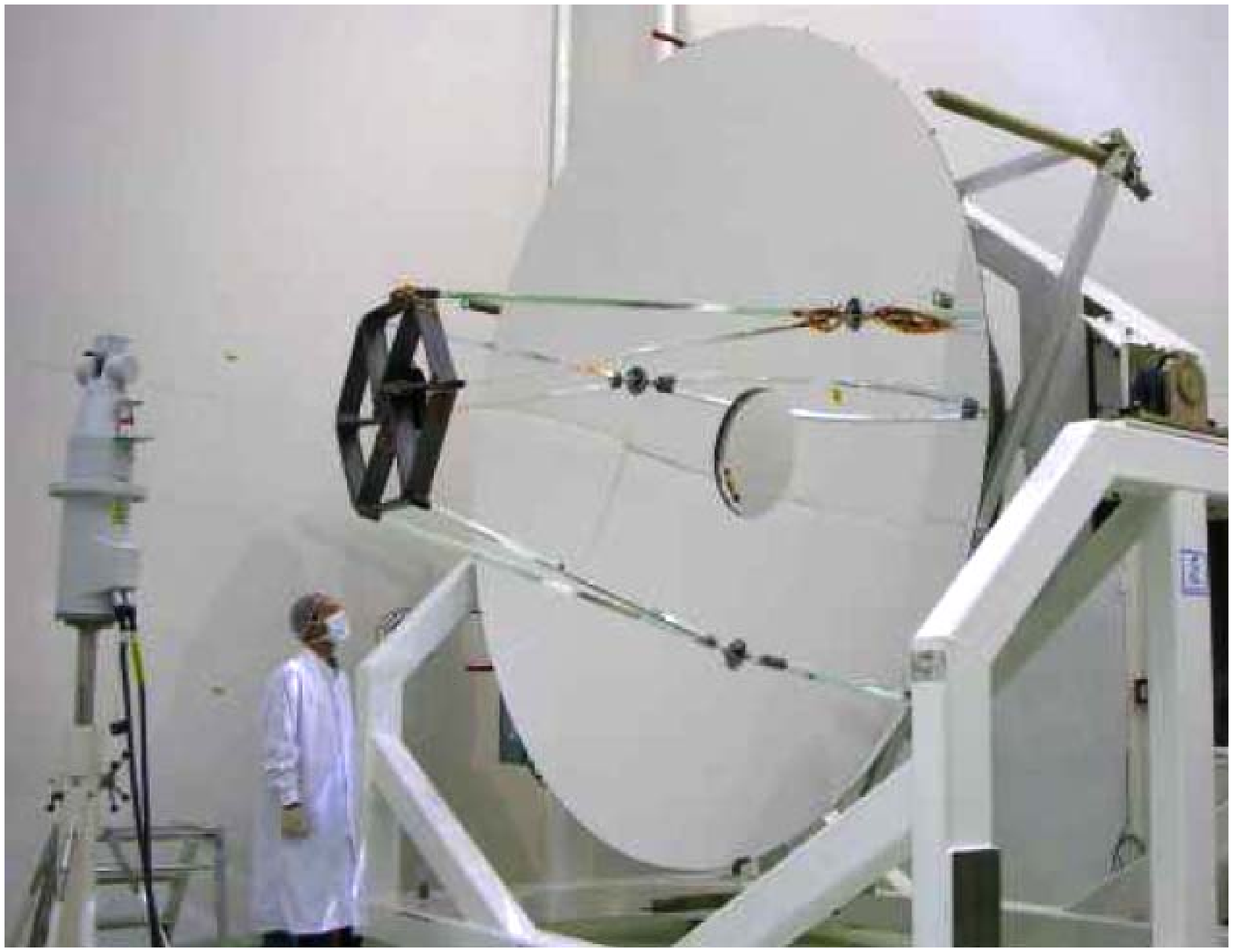}}
\caption{ESA {\sl Herschel} SiC telescope, 3.5m diameter mirror.}
\label{fig:bougoin4}
\end{figure}

\emph{Production of sintered SiC telescopes:} Since the end of the
1990's, {\sl Boostec} has developed space telescopes fully made of
SiC (mirrors, structure, focal plane), in close collaboration with
EADS Astrium. This experience can be summarized in 4 main steps:
\begin{itemize}
\item     1998-1999, Osiris Narrow Angle Camera which has been embarked on
            the ESA ROSETTA satellite, now traveling towards the
            Churyumov-Gerasimenko comet (Castel et al. 1999).
\item     2000-2001, {\sl Rocsat 2} telescope, a Cassegrain type remote sensing
            instrument now in operation for Taiwan (Uguen et al.
            2004).
\item     2001-2004, ESA {\sl Herschel} telescope to be launched in 2008
            (Fig.~\ref{fig:bougoin4}, Toulemont et al. 2004).
            Its 3.50-m parabola
            is undoubtedly the most challenging space mirror part that has
            ever been manufactured in the World.
\item     2003-2004: {\sl Aladin} instrument (Morancais et al. 2004). This 1.5m
            Lidar will be embarked on the {\sl Aeolus} ESA mission in 2008.
\end{itemize}

\begin{figure}
\resizebox{\hsize}{!} {\includegraphics[]{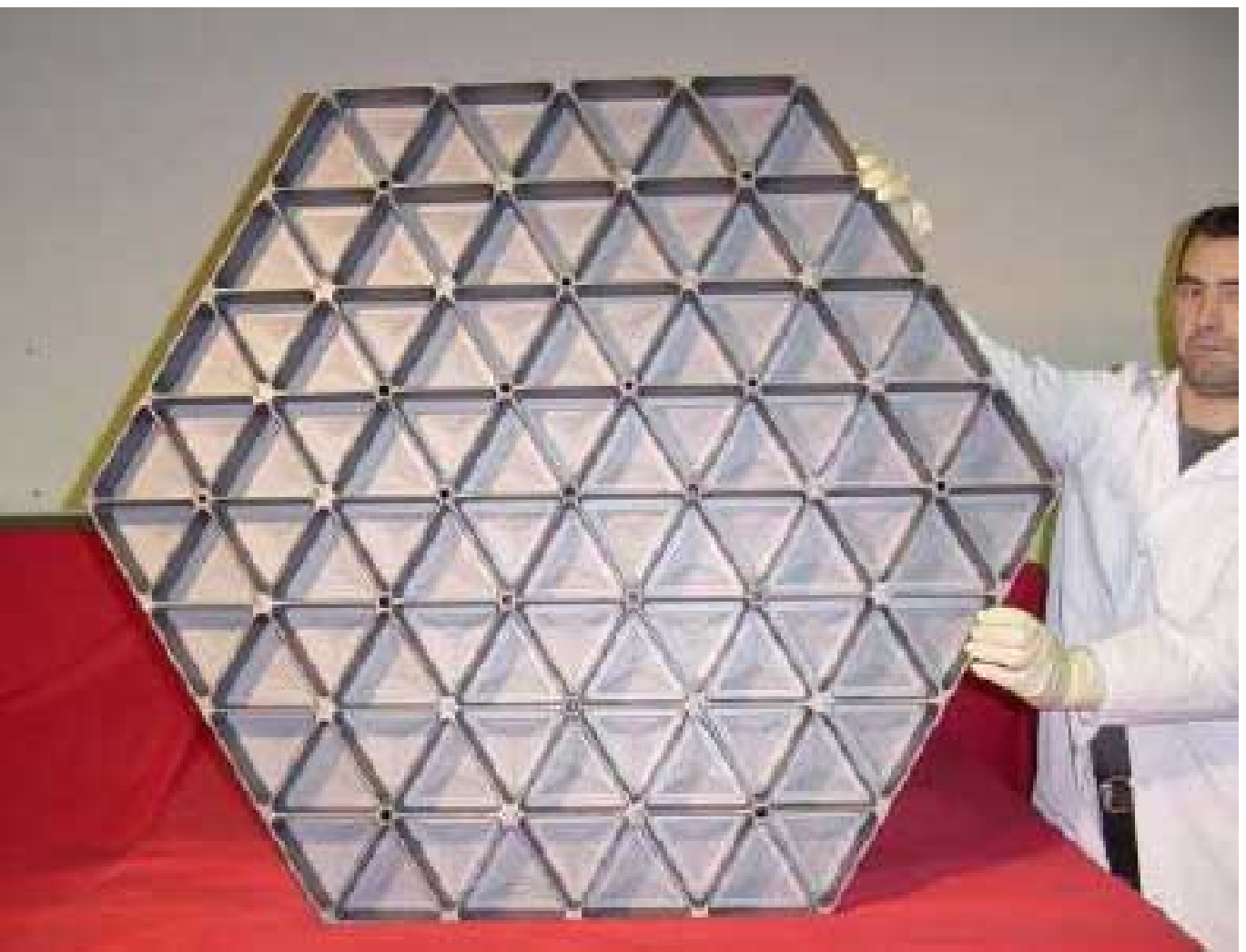}} \caption{OWL SiC
demonstrator, 1m flat to flat.} \label{fig:bougoin5}
\end{figure}

OWL-type segments have been designed with EADS Astrium assistance
(a) 1.00 meter flat to flat, (b) flat optical surface ready for
polishing or deposition of a polishable coating, (c) face sheet
thickness 5.5mm, (d) ribs thickness 4.0mm, (e) overall thickness
80mm (f) 19 possible points of interface. Their weight per unit area
is only 44 kg/m$^2$. Four segments demonstrators have been
successfully manufactured and delivered to ESO in 2003
(Fig.~\ref{fig:bougoin5}).

\subsubsection{Feasibility mirror segments for ESO's OWL}

The huge primary and secondary mirrors of OWL would be made of
thousands of hexagonal segments, the size of which should be between
1.3m and 2.3m. In 2001, {\sl Boostec} carried out feasibility and
cost studies for the fabrication of such segments with its SiC
technology. It demonstrated the easy feasibility of 20 to 30
segments per week. Such a fabrication would not bring any disruption
in the world production of sintered silicon carbide. The full set of
OWL segments blanks appeared also feasible within a competitive
price range.

A preliminary design analysis performed by ESO team (Dierickx et al.
2004) has shown that the savings on the total moving mass could be
around 5,000 tons (from 14,800 tons) when using SiC M1 segments
instead of glass-ceramics ones. Mass saving has positive impacts at
almost every level of the system design, in terms of performance,
safety and costs. Eliminating unnecessary structural masses reduces
flexures and stresses, implies less powerful drives, relaxes
requirements for the concrete foundations, reduces transient thermal
distortions, and simplifies integration (ESO OWL website,
www.eso.org/projects/owl/).

\subsubsection{Adaptive optics with SiC?}

ELTs will require adaptive optics for wave front error correction.
The reflective face and the back plate of such deformable mirrors
could be also advantageously made of SiC. The optical faces will be
large and thin shells, around 2mm thick, possibly segmented and then
off-axis. The back plates will be large and stiff structures,
including thousands of interfaces with the actuators. Even though no
technological obstacle is foreseen, such deformable mirrors in SiC
would undoubtedly require development work.

\subsection{High accuracy optical finishing through ion-beam
figuring}\label{S_IBF}

The focal plane instrumentation for the future ELTs will use large optical
surfaces made in different materials and probably with aspherical shapes. These
optics are generally difficult and expensive to manufacture within the tight
tolerances requested for astronomical instruments. A process that has been
proven to be effective in the high precision figuring of the optical surfaces
is the Ion Beam Figuring (IBF). This technique was originally developed by
Eastman Kodak Company in 1988 and is an excellent complement to conventional
figuring.

The optics are first polished conventionally and then the final figure is
milled by the IBF. The optics are inserted into a high-vacuum chamber facing
down. The IBF then directs a beam of ions upward to the glass that is hence
removed on a molecular level. The beam itself is translated across the optical
surface removing the errors leftover from the conventional polishing, leaving
behind a very smooth surface. The ion beam is produced by a Kaufman type ion
source using Argon gas. The shape of the beam (removal function) is ``Gaussian
like'' and its erosion rate (etching) is very constant in time for a given
setup of the parameters controlling the ion head, i.e. the electrical voltages,
the Argon flow, the distance ion head – mirror. The material is removed from
the optic by transfer of kinetic energy from the argon ions to the atoms of the
optical surface. The dwell time map and the movements of the ion source
necessary to remove the thickness of material are computed comparing the actual
surface with the target surface figure. The correct thickness of material is
then etched by rastering with the appropriate velocities the ion beam in front
of the optical surface.

This technique has the advantage that is deterministic (time saving), is a
non-contact method (good for lightweight optics) and has high removal rates
(50-100 nm/min). The main disadvantages are that it needs the vacuum (complex
technique), the substrate can heat during the figuring and if the quantity of
material removed is large (more than few microns) the surface micro-roughness
can increase depending from the material. This technique is of crucial
importance for figuring SiC optics because, due to the material hardness, it is
very difficult to reach a good figuring accuracy with classical methods. Of
course, other materials such as BK7, Zerodur and Quartz can be easily figured
as well.

\begin{figure}
\resizebox{\hsize}{!} {\includegraphics[]{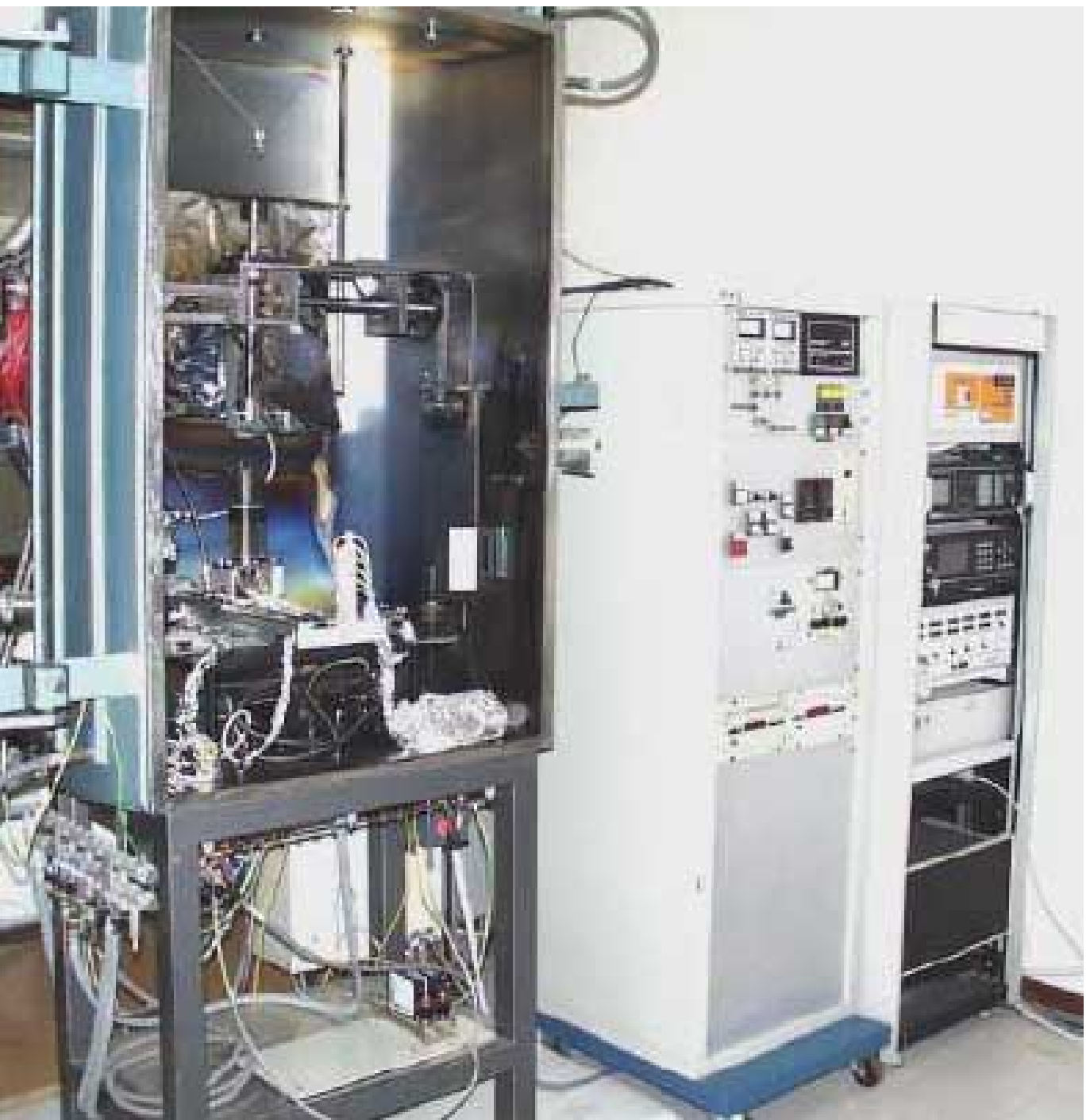}} \caption{The INAF-OABr
vacuum chamber.} \label{fig:ghigo1}
\end{figure}

An IBF facility has been recently developed at the INAF-Osservatorio
Astronomico di Brera (OABr) for the high precision machining of
optical surfaces. The facility (Fig.~\ref{fig:ghigo1}) consists of a
stainless steel vacuum chamber (1.4m height and 0.8m diameter),
suitable for figuring optics up to 500mm diameter. A two stage
mechanical pump is used for initial pump-down while the high vacuum
is obtained with a cryo-pump able also to take care of the small
volume of Argon gas used in the sputtering process. A ``Kaufman''
ion source is mounted on two carriages (x-y) with stepper motors.
A ``bridge'' is used to suspend the optic
above the source that can be moved in x-y to reach the point of the
surface to be figured.

The system is computer controlled and has been designed to be
autonomous and self-monitoring during figuring by using a
proprietary process control software. This software uses a time
matrix map indicating the dwell times required for each pixel of the
optical surface. This time matrix and the tool path necessary to
correct the errors are computed considering the actual surface error
map, the ion-beam removal function and the final target surface. The
software and the mathematical tools used to compute the time matrix
solution has been developed at INAF-OABr.

\begin{figure}
\resizebox{\hsize}{!} {\includegraphics[]{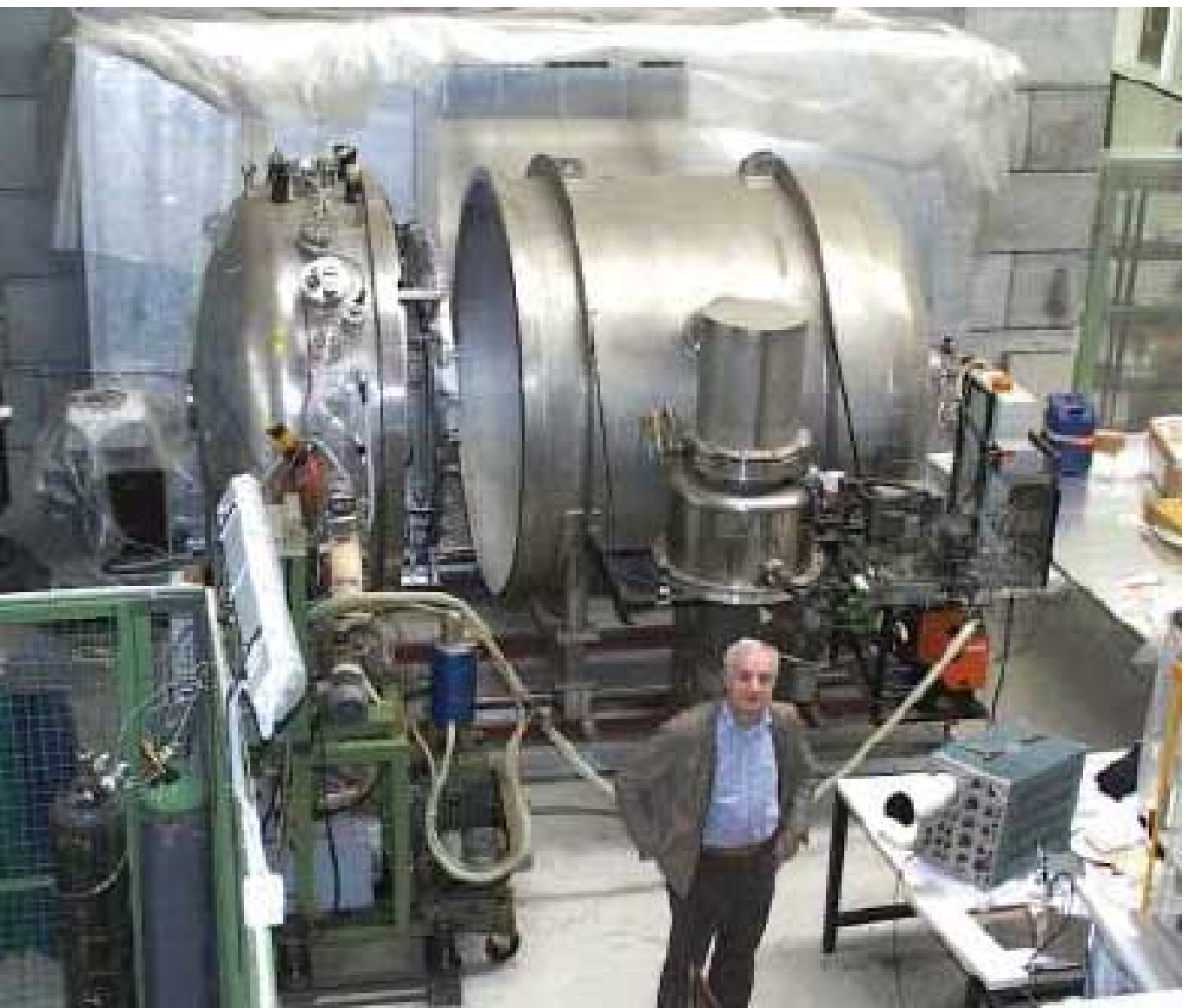}} \caption{New
IBF vacuum chamber at OABr.} \label{fig:ghigo5}
\end{figure}

An example of the performance that can be obtained with IBF are the
two 150-mm SiC demonstration mirrors for NIRSpec for the JWST. For
both mirrors it has been possible to obtain a precision of
$\lambda$/70 rms at 632.8 nm, confirming the high accuracy that can
be achieved with this technology. Currently, a new IBF facility is
under construction at OABr that will be able to figure optics up to
1.5m of diameter. Figure~\ref{fig:ghigo5} shows the vacuum chamber
with a dimension of 2m in diameter and 3m in length.

\subsection{Fused quartz and silica for large optics}

Crystal glasses as fused quartz and silica seem to be most
appropriate for large optical components in ELT instrument design,
due to their very high grade and reproducible properties.

\emph{Dimensions available:} Blanks with diameters up to 1m are now
available. Above this size, further development are required in
cooperation between the few producers in the world. Thickness will
vary according to diameter and weight restrictions. Indeed, overall
weight should be maintained below 500 kg.

\begin{figure}
\resizebox{\hsize}{!} {\includegraphics[]{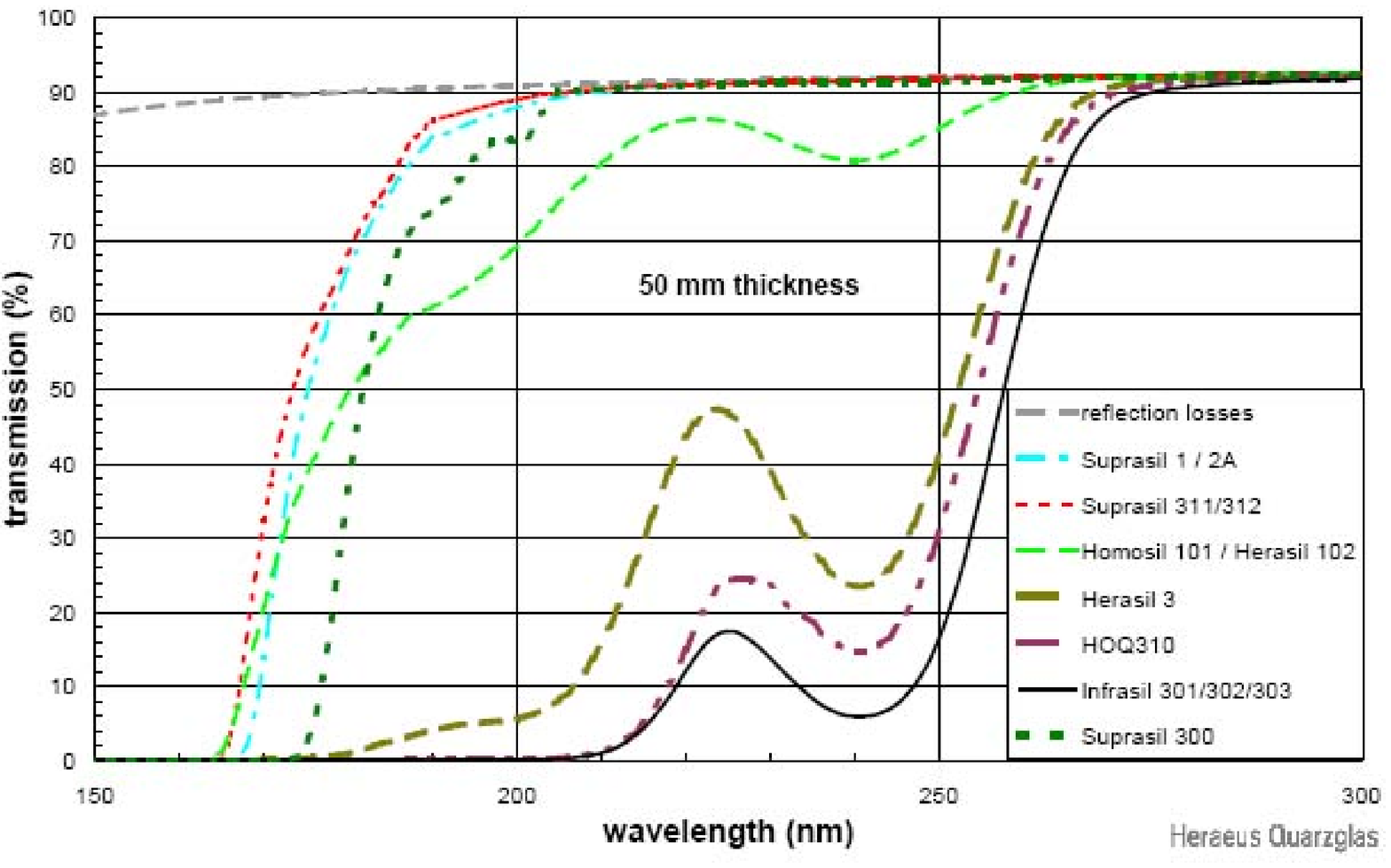}}
\resizebox{\hsize}{!} {\includegraphics[]{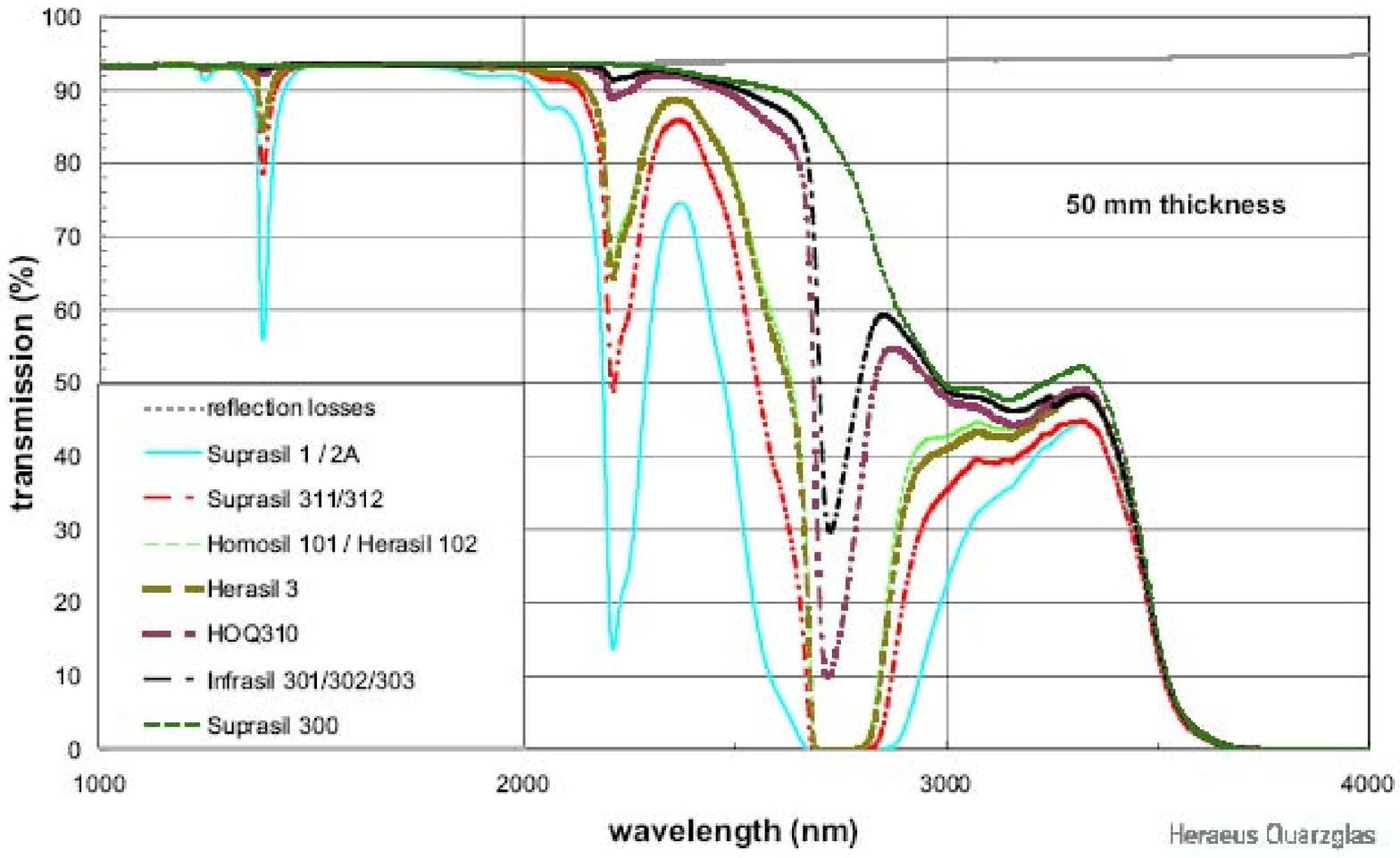}}
\caption{UV (top) and IR (bottom) transmission of fused Silica.}
\label{fig:takke1}
\end{figure}

\emph{Transmission:} In the infrared, transmission is influenced by
OH absorption bands, while in the ultraviolet by purity. See
Fig.~\ref{fig:takke1}. New materials are under development, such as
{\sl Heraeus} Suprasil-3001, with higher performances.

\emph{Birefringence:} For large pieces (diameter 350-1000mm) it
accounts for less than 5 nm/cm, while for smaller pieces ($<$350mm
diameter) values below 0.5 nm/cm are possible. These very small
numbers open manufacturing of very large optics.

\emph{Homogeneity:} For large pieces, homogeneity grade as low as
$\Delta$n=2\,10$^{-6}$ are possible. Moreover no high spatial
frequencies are present. For smaller pieces, homogeneity four times
smaller is feasible.

\emph{Bubbles and inclusions:} Conventional SiO$_2$ glasses are
affected by bubbles and solid inclusions at very different levels.
Very few glasses can be made with almost no bubbles and/or
inclusions. For example, {\sl Heraeus} Suprasil-312 is produced
without any imperfection.


\section{Smart Focal Planes}

There is a need to ensure that instruments for ELTs make best use of the high
information content of the focal plane, both for imaging and spectroscopy. The
OPTICON Joint Research Activity on Smart Focal Planes (Cunningham et al. 2005)
has developed a range of optical technologies to enable multi-object and
integral field spectroscopy, from robotic pick-off devices and reconfigurable
slit mechanisms to replicated image slicers.

\subsection{Pick-off technologies for multi-object ELT instruments}

Many of the scientific goals for future Extremely Large Telescopes,
including key fields such as the formation of galaxies at earlier
epochs or studies of the stellar population in neighbouring
galaxies, call for a multi-object capability from an instrument. It
has been proposed (Russell et al. 2004) that such an instrument may
be highly modular, comprising a number of identical imagers or
spectrographs observing in parallel. The number of sub-instruments
would depend on the limitations of the instrument conceptual design,
but may be as many as one per instrument. Current examples of
modular instruments include KMOS (3 spectrographs for 24 targets,
Sharples et al. 2004) and MUSE (24 spectrographs for 24 fields),
both being built for ESO's Very Large Telescopes. We call the
opto-mechanical means to pick-off areas of the focal plane means to
feed these sub-instruments ``Smart Focal Plane'' technology.

Pick-off technologies are an exciting field of opto-mechanical
development with several approaches of varying technical readiness
currently under development. The space densities, field sizes and
clustering of the science targets, coupled with instrument design
constraints, will drive the choice of multi-object technology. For
infrared astronomy on an ELT, the requirement to work at reduced
temperatures, often cryogenic, in order to minimize thermal
background radiation places significant constraints on the
technologies used to place pick-off mirrors and lenses in the focal
plane. Hydraulic mechanisms, for example, are obviously precluded.
The three techniques presented here; pick-off arms, Starbugs and a
``Planetary Positioner'' system have potential for use in a variety
of cryogenic instruments and highlight the issues and trade-offs
which will be required when developing Smart Focal Plane optics for
ELTs.

The UK Astronomy Technology Centre (UKATC) has designed the pick-off
technology for KMOS, a K-band multi-integral field unit spectrometer
that will use 24 pick-off arms to relay 24 fields to three
spectrometers.  Pick-off arms are attractive in that it is possible
to include a trombone style path length compensator into the arm to
ensure that the focal plane is reimaged at a fixed focal plane (in
the case of KMOS, onto fixed integral field units). Arms are also
attractive in that the position of each pick-off mirror can, with
suitable optical calibration, be ascertained by mechanical metrology
without the need for an independent optical metrology measurement
for each pick-off mirror placement. With 24 arms, it is obviously
important that the reliability of each of the mechanisms be
extremely high, which presents a challenge to the quality of the
cryo-mechanical engineering.  For an ELT instrument requiring of the
order of 100 pick-offs, the space requirements of an arm would
appear to preclude this technique as an option. The KMOS pick-off
arms are about to undergo cryogenic tests.

Starbugs is a new concept from AAO that uses piezo actuators to
place a miniature optical platform on an arbitrarily large focal
plane. The concept is being developed within the OPTICON Smart Focal
Planes project (McGrath et al. 2004). Starbugs include a magnet
within the platform which enables them to grip effectively to a
steel focal surface at any gravity vector.  On this platform a
variety of payloads such as pickoff mirrors or fibres may be
mounted. As each of the bug platforms can move simultaneously, a bug
speed as low as 1 mm/s can enable the pick-off observation pattern
to be reconfigured in as little as 5 minutes. Tests at temperature
down to -100~$^{\circ}$C have been conducted, with satisfactory,
though lower, operational speed recorded.  A destination arrival
accuracy near 10 $\mu$m with a loop cycle time of under 100ms has
been achieved. Operation in an ELT instrument would require the
introduction of an optical metrology system covering the entire
focal surface and control software not without some complexity.  A
further challenge for starbugs is that their present manifestation
requires power wires which trail behind each bug, and wireless
operation is currently being considered.

A further new technique which is scalable with regards object number
is the planetary positioner system; a pick-and-place system which
enables magnetic pick-off mirrors to be placed perpendicular to an
arbitrarily curved focal surface with the minimum of mechanisms.
Presently being built and tested at UKATC, ASTRON and CSEM within
the OPTICON Smart Focal Plane programme, the planetary positioner
system also has the benefit of an intrinsic measure of the position
of the mirrors.  However, though the time taken to position each
mirror can be relatively short, it will likely be necessary to
configure the mirrors "off-line" and rotate the focal surface into
place for the observation run in a similar way to that of ``2df''
(Colless 1999).

\subsection{Complex multi-sensory-motion systems}

Most concepts currently being conceived are based on a complex
layout comprising a large number of sensory-motion elements,
numbering in the hundreds, and eventually in the thousands, of
active components. This approach, beside its advantages in terms of
massive scientific capability, exponentially increasing with respect
to currently available instruments, offers as well a perspective for
escaping the a parallel exponential increase of cost by exploiting
more industrial technologies and production methods.

There are some technological developments which are recognized to be
potentially of use in next generation instrument concepts.

\emph{Mechanical slit masks:} for multi-object spectrographs.  This
mechanism creates a mask by placing two sets of bars. Multiple
objects can be selected in the available field-of-view, by
translating separate rectangular bars towards each other and
creating a rectangular opening (slit) on the focal plane of the
instrument.  Each bar has its own sensory-motion system, which
result in a quite complex multifold actuator mechanism.

\emph{Starbugs:} The starbug concept has been proposed since some
time (McGrath et al. 2004) as a next generation smart focal plane
system for astronomical instruments. From a technical standpoint, it
consists of a materialization of the focal plane as an opaque
surface, on which are placed and, in some version, may travel small
``starbugs'' that are optical pick-up elements which can have
various functions. Eventually, a starbug instrument would number
thousands of small opto-mechatronic systems. The current starbug
prototypes made by AAO are not autonomous, which limits greatly
their demonstration aspects. However, there exist a number of
projects for ``real'' micro-robot of comparable size. These
autonomous micro-robots are sensory-motors units with generally two
independent axes, hence functionally comparable to the what is
expected from a starbug, whether motor axes are used for mobility
or/and to move optical devices located on the robot.

\emph{MEMS (micro electro-mechanical systems):} have long be
envisaged as a key technology for astronomy, but this has not yet
materialized, beside the current slit mask development for JWST. The
figure shows a MEMS microlocomotion device developed at CSEM. The
MEMS has hexagonal shape with a footprint of about 0.5 cm$^2$.  It
comprises on its downside 150 ``legs'' which can move it in a
millipede manner with a speed of the order of 1 cm/sec.

\emph{3D technology:} may be proposed for the packaging of
electronics. This is based on the stacking of electronic components
(chips, plastic packages, sensors) placed on a film.  This solution
allows mounting the plastic components, ire-bond the chips on the
films and testing them, even making a burn-in before stacking
(depending on the applications).

\emph{Wireless power:} Power autonomy is an obvious issue for
starbugs. The precise quantification of such requirement will depend
on the operative duty-cycle. One option is to take advantage of
existing RFID (Radio Frequency Identification) technology components
and networking technology to
provide wireless power and data transmission from the starbugs
floor.


\section{The economics of astronomical instrument developments}

It is nowadays recognized that astronomy lives presently a golden
age.  Never before have such blooming of research instruments and
associated funds be available for astronomical developments.
Nonetheless, in spite of this relative abundance of financing and
technical resources, astronomical technology cannot yet pretend to
drive by itself the development of new technology. Even if the
budget of a new instrument for a large telescope is currently
planned in the range of 20 Meuro, such amounts are still one order
of magnitude less than a "small" space science mission, and two
orders less than an innovative space telescope such as JWST, not to
mention any serious military development. Therefore, because of
absolute economical resource limits, new concepts for astronomical
instruments remain tributary of existing developments and can only
take an essentially \emph{opportunistic approach} in which
technologies developed (and paid) on other contexts are utilized and
adapted. Even when such adaptation may and will generally require
significant engineering work, it has to be set on firm technological
ground and be provided with an already established production base.
The increasing size of projected instruments, associated to concepts
providing for large numbers of identical or similar components, even
reinforces this aspect.  If older, simpler instruments could be made
at institutes with special custom designs and fabrication, the
current concepts with multi-components would result exceedingly
expensive if each of these components requires a special design and
fabrication. Therefore astronomy developers need to become better
aware also of the economics underlying the development and
implementation of new technology suitable for their applications.

\emph{"Let us reduce costs by taking advantage of industrial components,
methods, etc."}: such a sentence is often heard among participants of an early
definition phase for new instruments and telescopes. Actually, this approach
tends to overlook many realities of industrial production.  First, the fact
that industrial products are not always cost-effective.

\begin{itemize}
\item An automobile is cheap, at least measured in terms of capability per kg.

\item But, the ASIMO humanoid robot (made by car manufacturer Honda) is very
expensive, even the much simpler AIBO (dog-like) is very expensive (per kg)
with respect to a car.
\end{itemize}

It is then quite evident that an astronomical instrument will by its very
specialized nature be more like ASIMO than a series production automobile.
Another key aspects to consider are some fundamental differences of the process
of development for industrial equipment with respect to an astronomical
instrument.

\begin{itemize}
\item Instrument: the customer (institute) is also the end user

\item \emph{Industry: the customer uses the object to produce other equipment or
goods}

\item Instrument: cost (price) is (should be) driven by technical requirements

\item \emph{Industry: cost (price) is driven by the added value which the costumer will
produce with the item.}

\item Instrument: technical requirements are initially set very severely, then
often relaxed, and ultimately astronomers are (almost) always happy with an
outcome often quite below initial requirements

\item \emph{Industry: the technical requirements are strictly derived and adapted to
the economic ones. If at some stage the economy fails its promise, the project
is dropped}
\end{itemize}

The fact that many future instruments will be based on multiple sensory-motion
systems, numbering in the hundreds, perhaps thousands, suggests that economy of
scale can play a key role in the costing of such instruments. Yet, economy of
scale is a broad concept, not to be simply treated to mean that the cost of
making multiple copies of a successfully demonstrated prototype decreases
manifold. Establishing an economy of scale for a given component may take many
steps and levels. Typically each step "saving" on the unit production cost
requires investment in any of many domains:
\begin{itemize}
\item Production facilities.
\item Organization and personnel culture.
\item Packaging (e.g. wiring).
\item Control, inspection (quality, etc.).
\item Assembly.
\end{itemize}

Furthermore, maturity (i.e. number of product generations) is also an essential
aspect of a profitable economy of scale.

Clearly, industry usually active in the business of industrial
components cannot view and price its product to an institute in the
same way it does for an industrial customer. Thus it should be up to
the instrument institute to provide industry with different
incentives that will contribute to common interests. Instrument
developing institutes should take a mutually constructive approach
in their relations with industrial partners.

 From the production point of view, one still does not
see that the optical transmitting elements are treated with the attention
necessary to provide the performance needed for the telescopes of the next
generation. There is a need for close cooperation between the glass
manufacturer, the polishing companies and the optical designers of the
telescopes to prevent bad surprises.


\section{Summary and conclusions}

ELT instrumentation will require significant developments in the
area of optical design, material production, optical manufacturing,
metrology, assembling, and opto-mechanics. A lot of efforts are
underway to increase present capabilities and performances of
classical optical components such as mirrors and lenses, mostly in
order to obtain larger sizes, better homogeneity, higher surface
quality a.s.o.. New materials and technologies are under study to
find solutions for the next generation ground telescopes and their
instruments.

In this paper, we showed how classical and new-technology optics can
be employed and where new developments are needed. Here we summarize
current results in different fields of optical engineering.

\begin{itemize}
\item Dimensions of the transmissive optical elements must be kept as small as
possible. With large lenses this holds especially for their
thickness. Thickness determines the thermal inertia of the lens,
which is the cause for many problems in production and in operation
of optical elements. Due to the quadratic influence of temperature
differences in the glass volume, even small thickness reductions
will help.

\item Internal imperfections will lead to stray light. Or they may become
visible if they are close to an intermediate image plane.

\item Segmented filters of large size can be constructed today. The edges of the elements
of a large mosaic filter may be treated like striae. If this is
taken into account in the design of the total optical system,
locations may be found where imperfections will not disturb or where
they can be tolerated or corrected in the subsequent image
processing.

\item Reflective large optics should be used often.

\item Optical transmitting elements are essential for the function of large
telescopes. All of the immense efforts in designing and constructing
of ELT mirrors and instruments may be in vain, when the intermediate
optics fail.

\item Volume phase holographics gratings are a mature technology; not
just for regular size spectrometers but also ready for extremely
large instruments. A monolithic grating of 380~mm of diameter
exists, and size can be further enlarged by using the mosaic
technique. These gratings stand cryogenic temperatures and work
properly in these conditions considering both efficiency and
wavefront errors. Moreover, the diffracted wavefront can be
corrected by polishing the blank after the grating has been
encapsulated. All these aspects make VPHGs attractive as dispersive
element for ELT spectrometers.

\item Diarylethene photochromic materials are useful and flexible in astronomical
instrumentation. Exploiting their change of transparency in the
visible wavelength range, rewritable focal plane masks were built,
characterized, and successfully tested. The performances are good in
terms of writing slits. The contrast between the slits and the
opaque background can be increased by using filters well tailored
for the specific photochromic material.

\item Photochromic polymers were synthesized to make self-standing films with
large modulation of the refractive index. These films could be the
base of highly efficient, near-infrared VPHGs working in the J-, K-,
and L-band. Present results suggest that photochromic polyesters are
promising materials for VPHGs to be used in next generation
astronomical spectrographs.

\item With an ion beam figuring facility it will be possible to make thin
adaptive mirrors, segments for large telescopes, large light-weight
optics for space applications and, of course, focal plane optics for
large instrumentation. In conclusion, it is clear that ion beam
figuring is a technique that will be useful for the manufacturing of
complex optical surfaces that will be necessary for ELTs.

\item Sintered silicon carbide (SSiC) shows very attractive characteristics for
space or ground telescopes. Its specific stiffness could allow
saving of moving mass. A lot of technological developments and
qualifications were successfully applied to large space telescopes,
making this technology ready for large ELTs optics manufacturing.

\item Free-form optics enlarge the parameter space available to optical
designer to meet instrumental requirements. It even allows to
correct for aberrated instruments when such optical elements are
inserted along the light path in an existing instrument.

\item Metrology plays a fundamental role for ELT instruments, both during the construction phase
and the operational phase. Wavefront sensing, adaptive optics,
mirror edge sensing and segmented actuators offer additional degree
of freedom to relax manufacturing tolerances, both for telescopes
and their instruments.

\item Aspherical surfaces are always challenging and very need careful
preliminary analysis in order to be successfully employed. No
recipes exist but test set-ups need to be searched for.

\end{itemize}

\acknowledgements

OPTICON has received research funding from the European Community's
Sixth Framework Programme under contract number RII3-CT-001566.

\end{document}